\documentclass[prl,aps,twocolumn,showpacs,preprintnumbers,amsmath,amssymb]{revtex4-1}

\usepackage{graphicx}
\usepackage{dcolumn}
\usepackage{bm}
\usepackage{siunitx}
\usepackage[outdir=./]{epstopdf}
\usepackage{eucal}
\usepackage{color}
\usepackage{graphicx}
\usepackage{subcaption}
\usepackage{xcolor}
\usepackage{amsmath} 

\begin{document}
	\title{Free carrier induced ferroelectricity in layered perovskites}
	\author{Shutong Li}
	\author{Turan Birol}%
	\email{tbirol@umn.edu}
	\affiliation{Department of Chemical Engineering and Materials Science, University of Minnesota}

\begin{abstract}
Doping ferroelectrics with carriers is often detrimental to polarization. This makes the design and discovery of metals that undergo a ferroelectric-like transition challenging. In this letter, we show from first principles that the oxygen octahedral rotations in perovskites are often enhanced by electron doping, and this can be used as a means to strengthen the structural polarization in certain hybrid-improper ferroelectrics -- compounds in which the polarization is not stabilized by the long range Coulomb interactions but is instead induced by a trilinear coupling to octahedral rotations. 
We use this design strategy to predict a cation ordered Ruddlesden-Popper compound that can be driven into a metallic ferroelectric-like phase via electrolyte gating. 
\end{abstract}
	
	\date{\today}
	\maketitle 

Ferroelectrics, insulators with a spontaneous and switchable electric polarization, are promising for a wide range of applications and pose a number of fundamental questions \cite{Dai2020,Rabe2007,Martin2017,Scott2007}. While ferroelectricity is observed in a wide range of material classes and can be driven by a variety of mechanisms, the most studied ferroelectrics are transition metal oxides, such as BaTiO$_3$, where the emergence of a polar order parameter is due to a crystal structural distortion driven by the interatomic hybridization and long range Coulomb interactions \cite{Ghosez1996, Cohen1992}. 
Because of the role of the long range interactions in driving the polar structural distortion, introduction of free charge carriers to ferroelectrics not only screens the ferroelectric polarization, but it also suppresses the structural distortion often \cite{Wang2012}. 

While `structurally polar metals' (metals with a polar point group) are rather common, `ferroelectric metals' (metals that undergo a phase transition from a centrosymmetric to a polar crystal structure \cite{benedek2016ferroelectric}) are rather rare. It took almost 50 years after the possibility of a ferroelectric-like transition in a metal was first raised \cite{Anderson1965} for the unambiguous experimental observation of such a transition in LiOsO$_3$ \cite{Shi2013}.  The first observation of polarization switching in a ferroelectric (semi-)metal is even more recent \cite{Fei2018, sharma2019room}. 
The interest in polar and ferroelectric-like metals is continuing to increase in both bulk and heterostructures \cite{Meng2019, Xiang2014, LoVecchio2016,Liu2015,Narayan2020,Wang2021,Jin2019, Shan2020,Yimer2020,Ghosh2017,Zabalo2021,Xiao2020} 
and they continue to promise both a fertile playground for interesting emergent phenomena (including, but not limited to mixed singlet-triplet superconductivity \cite{Salmani-Rezaie2020} and novel optical effects\cite{Mineev2010}), and immediate relevance to applications as polar electrodes \cite{Puggioni2018}. 

Emergence of polarization in (Sr,Ca)Ru$_2$O$_6$, Ca$_3$Ru$_2$O$_7$, and ultra-thin NdNiO$_3$ films have been studied in detail \cite{Puggioni2014,Kim2016, Lei2018}; and it was shown that the polarization in these materials is robust against metalicity because the polar displacements are driven by their coupling to zone-boundary phonon modes and are mainly decoupled with the electrons around Fermi level.
`Metallized ferroelectrics' (insulating ferroelectrics that are doped to introduce charge carriers) are also studied intensively, and the effects of free carriers on the polarization and polar instabilities are analyzed recently introducing ideas such as metascreening \cite{Zhao2018}, and elucidating the trends in the second-order Jahn-Teller effect under carrier doping \cite{Hickox-Young2020}. 
Barring a volume expansion, the most common effect of charge doping in proper ferroelectrics is the suppression of the ferroelectric polarization; for example, $\sim$0.11 electrons per  formula unit is sufficient to completely suppress the polarization in BaTiO$_3$ and render it centrosymmetric \cite{kolodiazhnyi2010persistence, Wang2012, Xia2019}.

In this letter, we show that a particular group of ferroelectrics, the A$_3$Sn$_2$O$_7$ hybrid-improper ferroelectrics (HIFs) \cite{Benedek2011, Wang2017, Yoshida2018, Lu2019, Chen2020} behave differently, and their structural polarization is strongly enhanced by the free electrons introduced by chemical doping or electrostatic gating. This is related to an increase in the oxygen octahedral rotation angles induced by the added electrons in the parent perovskite compounds, which in turn leads to a larger structural polarization in these layered perovskite Ruddlesden-Popper (RP) phases \cite{Ruddlesden1958}. We also show that it is possible to exploit this mechanism to obtain free carrier \textit{induced} polarization, in other words, design a material that develops a ferroelectric-like structural instability when free electrons are introduced via, for example, electrostatic or electrolyte gating.

\begin{figure}
\includegraphics[width=0.45\textwidth]{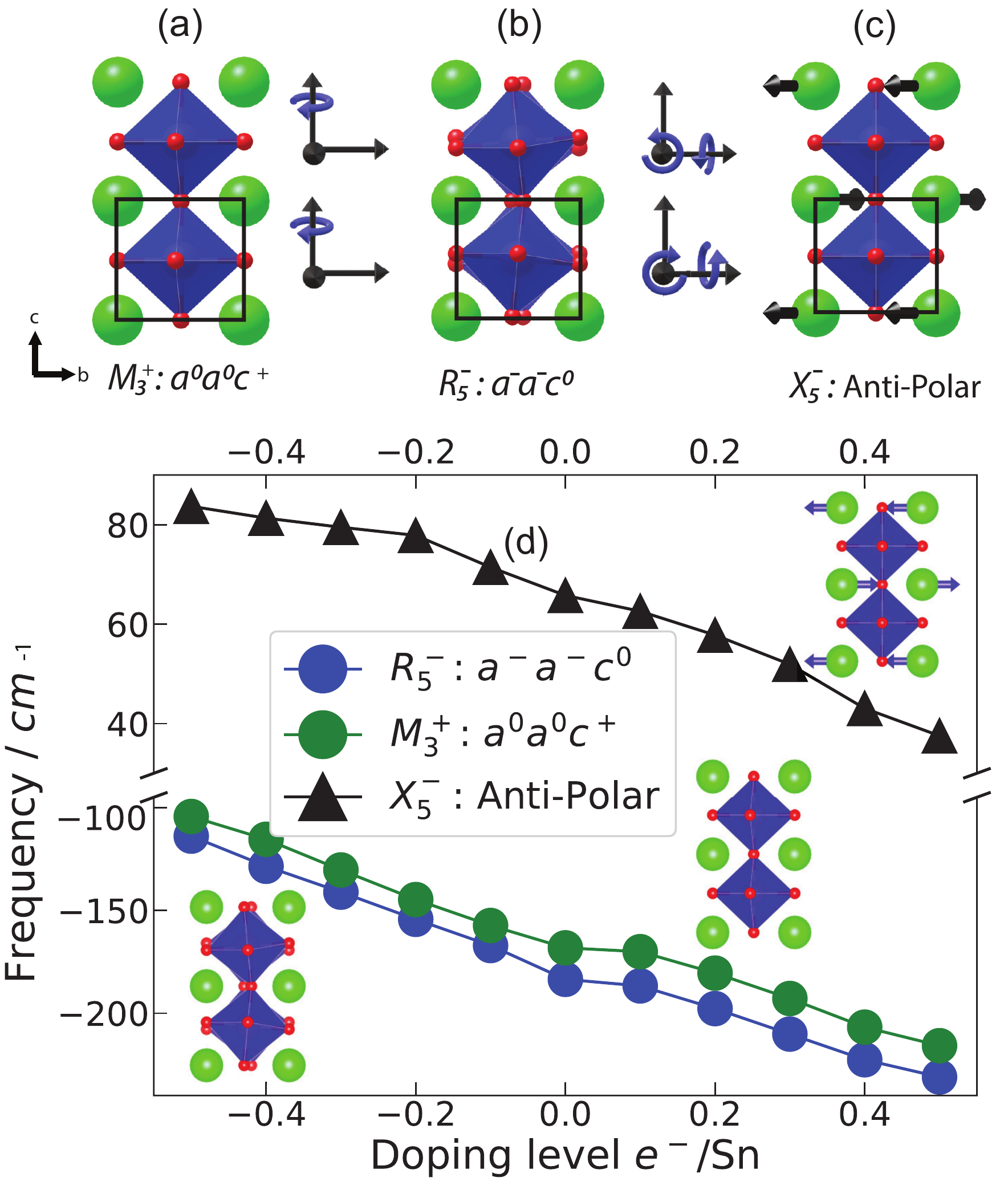}
\caption{(a)-(c)The three normal modes in Eq.~\ref{equ:freeenergy} that are relevant to the $Pnma$ phase of perovskites and (d) (d) The phonon frequencies of cubic ($Pm\bar{3}m$) SrSnO$_3$ under doping and fixed volume. The green and red spheres represent the A-site and oxygen ions respectively, and the B-site atoms are in the center of the blue octahedra. (a) In-phase rotation around the c-axis ($a^0a^0c^+$ in Glazer notation). (b) Out-of phase rotation around the ab-axis $a^-a^-c^0$. (c) The anti-polar displacement in the ab-plane, where the irrep direction is $X_5^-(a,a; 0,0; 0,0)$. With increasing number of electrons, the unstable rotation modes get more unstable, and the anti-polar $X_5^-$ mode gets softened (but remains stable).}
\label{fig:distortion_1}
\end{figure}

\textit{Oxygen Octahedral Rotations in Perovskites --- }
We start by reviewing the symmetry and octahedral rotations in ABO$_3$ perovskites. Most perovskite oxides have the orthorhombic space group of $Pnma$ at low temperature \cite{lufaso2001}. The atomic displacements that lead to the $Pnma$ symmetry can be expanded in terms of the irreducible representations (irreps) of the reference space group $Pm\bar{3}m$ \cite{Aroyo1998, Campbell2006}. 
The $Pnma$ structure has multiple nonzero strains ($\Gamma_1^+$, $\Gamma_3^+$, $\Gamma_5^+$) and atomic displacements ($R_4^-$, $R_5^-$, $X_5^-$, $M_2^+$, and $M_3^+$). 
Most of these distortions are `secondary': they are nonzero only because of couplings with other, `primary' distortions. The $Pnma$ structure can be obtained by a combination of only two primary irreps ($R_5^-$ and $M_3^+$) which correspond to the out-of-phase and in-phase oxygen octahedral rotations shown in Fig.~\ref{fig:distortion_1}a-b \cite{Woodward1997a}. Both of these space irreps are 3 dimensional, where the 3 orthogonal directions correspond to octahedral rotations around the three cubic axes. The phonons corresponding to both of these octahedral rotations are unstable in the cubic reference structure of $Pnma$ perovskites. The $Pnma$ structure ($a^-a^-c^+$ in the Glazer notation) has out-of-phase rotations around [110] and in-phase rotations around [001], which is equivalent to order parameter directions $R_5^-(a,a,0)$ and $M_3^+(0,0,a)$. 
These two modes, which we henceforth refer to as $R$ and $M$ for brevity, couple with the $X_5^-(a,a; 0,0; 0,0)$ mode (referred to as $X$ for brevity) at the trilinear order. Hence, the Landau free energy up to third order is 
\begin{equation}
\mathcal{F}=\alpha_R R^2 +\alpha_M M^2 +\alpha_X X^2 + \gamma R\cdot M\cdot X 
\label{equ:freeenergy}
\end{equation}
The X mode corresponds to an out-of-phase displacement of the A-site cations as showed in Fig.~\ref{fig:distortion_1}c, and is typically stable, but it has a nonzero amplitude $X=\gamma R M/2\alpha$ in the low temperature structure. 
$X$ can be referred to as a `hybrid-improper' order parameter, because it is induced in the ground state by a combination of two primary order parameters. 

In heterostructures where translational symmetry is broken by layered cation ordering, or in layered perovskites (RPs), modes that give rise to transverse out-of-phase displacements of the A site (related to the $X_5^-$ in perovskites) attain a polar character, and are responsible of the hybrid-improper ferroelectricity \cite{Benedek2011, Benedek2015, Mulder2013}. For this reason, understanding the behavior of this mode in bulk perovskites is essential for understanding the polarization trends in HIFs. 
As an example $Pnma$ perovskite system, we consider SrSnO$_3$. While SrSnO$_3$ is orthorhombic at room temperature, its Goldschmidt tolerance factor $t = \frac{R_{Sr}+R_O}{\sqrt{2}(R_{Sn}+R_O)}=0.96$ is close enough to 1 so that it undergoes a series of phase transitions to the cubic phase above 1295 Kelvin and its structural ground state can be modified by biaxial strain \cite{GLERUP2005507, Wang2018}. 
In Fig.~\ref{fig:distortion_1}(d), we show the phonon frequencies for the $R$, $M$, and the $X$ modes as a function of doping from first principles DFT calculations. (The technical details of the calculations are discussed in the supplement \cite{Supplement}.) The phonon frequencies are proportional to the square root of the $\alpha$ coefficients in Eq.~\ref{equ:freeenergy}, and can be used to study the instabilities. Unstable modes have imaginary frequencies, which are plotted as negative numbers. We simulate the effect of free carriers in this nominally insulating compound by changing the total number of electrons in the calculation, while keeping the system neutral by adding a homogeneous background charge. Unlike chemical substitution, this approach does not introduce any steric differences or disorder into the system. In this respect, it is a better representation of electrostatic or electrolyte gated systems rather than chemical doping. We consider a wide range of carrier doping up to 0.5 electrons per Sn atom, which is larger than the typical concentrations experimentally achievable \cite{Yuan2009}. We keep the unit cell volume fixed in order to separate out the volume expansion effects. The volume expansion does not modify the trends we report significantly \cite{Supplement}.

\begin{figure}
\centering
\includegraphics[width=0.5\textwidth]{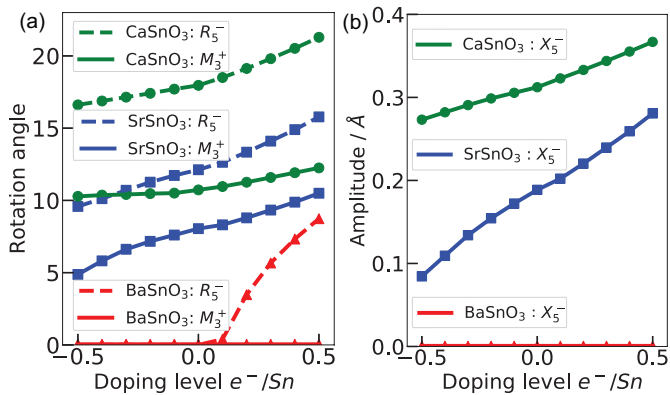}
\caption{Results of DFT structure optimizations under fixed volume. Both (a) the octahedral rotation angles (b) the anti-polar mode amplitudes increase with increasing number of electrons, as expected from the phonon frequencies in Fig.~\ref{fig:distortion_1}(d). }
\label{fig:Sn_relaxed}
\end{figure}

\begin{figure}
\centering
\includegraphics[width=0.5\textwidth]{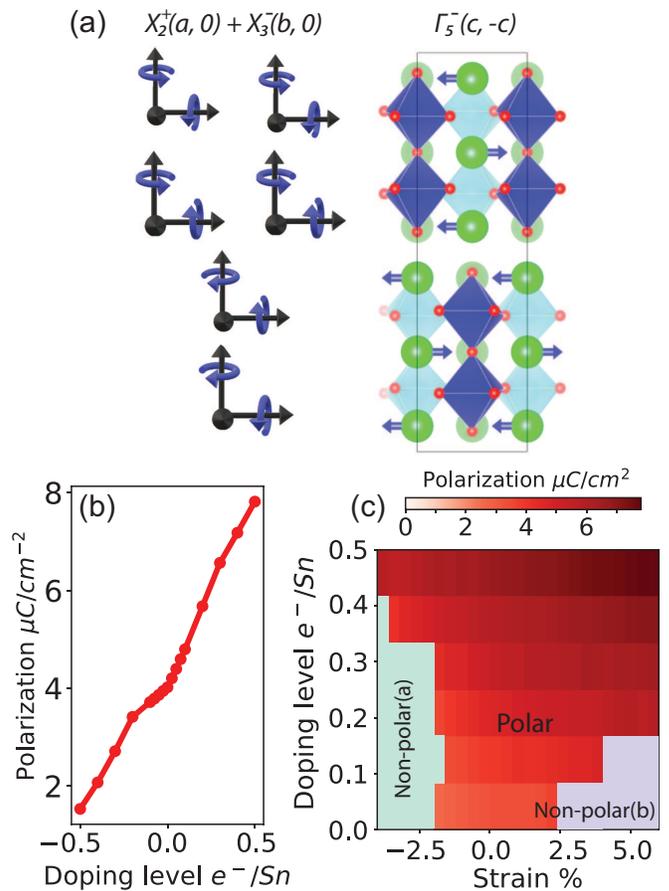}
\\
\caption{(a) The structure of Sr$_3$Sn$_2$O$_7$ includes three structural distortion modes with respect reference $I4/mmm$ structure: Two oxygen octahedral rotation modes ($X_2^+$ and $X_3^-$) and a polar mode ($\Gamma_5^-$). (b) The structural polarization strength of Sr$_3$Sn$_2$O$_7$ as a function of doping level. The structural polarization is as the sum of the products of the nominal charge and polar displacements of ions \cite{Supplement}. (c) The doping -- strain phase diagram of Sr$_3$Sn$_2$O$_7$. The non-polar(a) and non-polar(b) phases have $Aeaa$  and $P4_2/mnm$ space groups respectively \cite{Supplement}.}
\label{fig:DOS}
\end{figure}

The results in Fig.~\ref{fig:distortion_1}d show that both the rotation modes $R_5^-$ and $M_3^+$, which have imaginary frequencies in the undoped compound, become more unstable with the introduction of free electrons, in other words, $\alpha_R$ and $\alpha_M$ become more negative with added electrons. Similarly, the frequency of the stable $X_5^-$ mode decreases with increasing electron concentration, and so $\alpha_X$ becomes smaller.
The trilinear coupling $\gamma$ doesn't change significantly under doping, and the changes in the higher order coefficients are qualitatively insignificant \cite{Supplement}. 
As a result, the softening of $X_5^-$ and the strengthening of the $R_5^-$ and $M_3^+$ instabilities under electron doping lead to larger rotation angles and antipolar amplitudes as shown in Fig.~\ref{fig:Sn_relaxed}. This trend is observed not only in $Pnma$ perovskites SrSnO$_3$ and CaSnO$_3$, but also in cubic perovskites like BaZrO$_3$, which develops a $R_5^-$ instability when electron doped \cite{Supplement}. 
A similar enhancement of octahedral rotations was predicted by DFT in SmNiO$_3$ \cite{Kotiuga2019}; and both DFT and X-ray diffraction points to enhanced octahedral rotations in photodoped EuTiO$_3$ \cite{Porer2019}. 

This effect of carriers on octahedral rotations can be explained by considering the densities of states. The valence bands in stannates consist of oxygen-p bands, whereas the the conduction band is formed by Sn-s \cite{Mizoguchi2004}. The added electrons fill states with Sn-s character, and the valence of Sn$^{4+}$ becomes Sn$^{+4-\delta}$. This decreases the Sn-O electrostatic attraction, decreases the Sn-O hybridization, and increases the ionic radius of Sn. This reduces the tolerance factor $t$. 
Added holes, on the other hand, occupy the O$^{2-}$ anions and make them O$^{-2+\delta}$. This reduces the attraction between the A site cation (Ba, Sr, or Ca) and oxygens, which is the driving force of rotational instabilities. Hence, rotation modes become less unstable.

\textit{Hybrid-Improper Ferroelectrics --- } 
We now move on to A$_3$B$_2$O$_7$ HIFs, and consider Sr$_3$Sn$_2$O$_7$ as an example. Sr$_3$Sn$_2$O$_7$ is an $n=2$ RP compound, which can be considered as a layered perovskite with an extra SrO layer after every pair of SrSnO$_3$ bilayers. It is experimentally verified to be a ferroelectric \cite{Wang2017, Yoshida2018}, and its polarization is induced through the hybrid-improper mechanism which involves the trilinear coupling between the polar mode ($\Gamma_5^-$, which we denote as $P$) and two octahedral rotation modes ($X_3^-$ and $X_2^+$, which we denote as $Q_1$ and $Q_2$) shown in Fig.~\ref{fig:DOS}a. These modes are the counterparts of the antipolar A-site displacement $X_5^-$ mode, and the octahedral rotation modes $R_5^-$ and $M_3^+$ in bulk perovskites. Two crucial differences between the A$_3$B$_2$O$_7$ RP and the ABO$_3$ perovskite structures are (i) in the smaller Brillouin zone of the RP structure, both the octahedral rotation modes $Q_1$ and $Q_2$ have the same wavevector, and hence can couple to zone center modes at the trilinear order, and (ii) the out-of-phase A-site displacement is now a polar $\Gamma$ mode because the dipole moments induced by the symmetry inequivalent A-sites don't cancel.
The shortest free energy that explains the polarization up to third order is 
\begin{equation}
\mathcal{F}=\alpha_{1} Q_1^2 +\alpha_{2} Q_2^2 +\alpha_P P^2 + \gamma Q_1 Q_2 P 
\label{equ:freeenergyRP}
\end{equation}
The trilinear coupling $\gamma$ between the unstable $Q_1$ and $Q_2$ rotations with $\alpha_{1,2}<0$ and the stable polar mode $P$ with $\alpha_P>0$ gives rise to a nonzero polarization $P=\gamma Q_1 Q_2/2\alpha_P$ in the groundstate. 

In order to elucidate the change in the structural polarization in Sr$_3$Sn$_2$O$_7$ when free carriers are introduced, we optimize the crystal structure again with different numbers of added electrons or holes. The results in Fig.~\ref{fig:DOS}b show that added electrons increase the polarization, similar to the increased antipolar $X$ mode amplitude in SrSnO$_3$. This can be explained by the fact that the mechanism that leads to enhancement of octahedral rotations in the electron doped SrSnO$_3$ is essentially a local mechanism that also applies to Sr$_3$Sn$_2$O$_7$, which also has a similar DOS with Sn-s bands on the conduction band. Filling the conduction band increases the effective ionic radius of the Sn ions, which in turn increases the amplitude of $Q_1$ and $Q_2$ octahedral rotations, and hence enhance the polarization $P$. 
    In the HIF Ca$_3$Ru$_2$O$_7$ or the proper geometric ferroelectric-like LiOsO$_3$, the polarization is persistent against free carriers because of the absence of significant coupling between the electronic states near the Fermi level and the unstable phonons \cite{Laurita2019, Puggioni2014}. In Sr$_3$Sn$_2$O$_7$, there is a strong effect of the conduction band occupation on the lattice instabilities, which is not reported in these other metallic ferroelectric-like compounds.

The enhanced rotations also expand the biaxial strain range where Sr$_3$Sn$_2$O$_7$ is structurally polar. In Fig.~\ref{fig:DOS}c, we show the strain -- doping phase diagram of Sr$_3$Sn$_2$O$_7$, calculated by fixing the in-plane lattice parameters and relaxing the out-of-plane one to simulate the boundary conditions on a thin film lattice matched to a substrate. Insulating, undoped Sr$_3$Sn$_2$O$_7$ is known to undergo a transition to a non-polar phase above $\sim\mp 2\%$ biaxial strain \cite{Li2020} like many other compounds \cite{Lu2016}. Fig.~\ref{fig:DOS}c shows that not only doping enhances polarization at fixed volume, but it also stabilizes the polar phase at wider strain ranges. 
The polar/non-polar transition induced by epitaxial strain is driven by the disappearance of one of two rotation modes in the polar phase \cite{Li2020}. The free electrons increase the stability of both rotation modes which make this phase transition occur at a higher strain value. 

\begin{figure}
\centering
\includegraphics[width=0.45\textwidth]{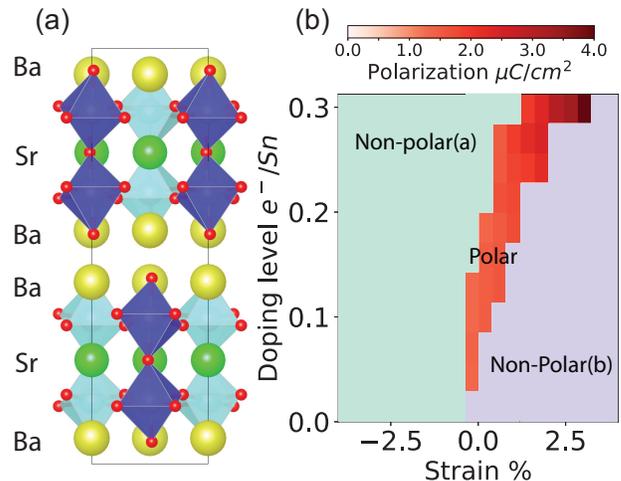}
\caption{(a) The structure and (b) a doping-strain phase diagram of Ba$_2$SrSn$_2$O$_7$. Yellow spheres represent Ba atoms. The non-polar(a) and non-polar(b) phases have $Aeaa$ and $P4_2/mnm$ symmetrıes respectively. }
\label{fig:Ba2SrSn2O7}
\end{figure}

\textit{Ferroelectric-like transition induced by free electrons --- } 
The strong effect of free electrons on stabilizing a metallic ferroelectric-like phase in Sr$_3$Sn$_2$O$_7$ leads to the question \textit{whether it is possible to drive a centrosymmetric compound to a polar phase by doping it with free electrons} without the help of biaxial strain. 
We scanned a number of A$_3$B$_2$O$_7$ oxides, but could not find an example that undergoes a polar phase transition for dopings up to 0.5 e$^-$ per B site cation, which is already beyond what is experimentally achievable via methods such as electrostatic gating. In order to \textit{design} a material which is closer to a structural phase transition than Sr$_3$Sn$_2$O$_7$, we turn to \textit{targeted chemical pressure}, which involves selectively substituting part of Sr ions with larger Ba cations \cite{Dawley2020}. While it is not always possible to order same charge cations in bulk, molecular beam epitaxy has been successfully used to obtain targeted chemical pressure in other RP phases (SrTiO$_3$)$_n$(BaTiO$_3$)$_m$SrO \cite{Dawley2020}.
In Sr$_3$Sn$_2$O$_7$ ceramics, up to 10\% of Ba ions are reported to preferentially substitute inequivalent Sr sites, however, the ordering tendencies depend sensitively on changes in the substitution amount \cite{Chen2020}. 
We consider a structure where the 2/3 of Sr cations are substituted with Ba to form Ba$_2$SrSn$_2$O$_7$, 
where the Ba cations are on the double-rocksalt layers of the RP structure, as shown in Fig.~\ref{fig:Ba2SrSn2O7}a. While this structure is not energetically the most stable one \cite{Supplement}, it may in principle be synthesized via layer-by-layer growth. 
The lowest energy structure of Ba$_2$SrSn$_2$O$_7$ is centrosymmetric when undoped and strain-relaxed, but introducing electrons to the conduction band leads to a transition to a polar structure with space group $Cmc2_1$ (Fig.~\ref{fig:Ba2SrSn2O7}b). Thus, Ba$_2$SrSn$_2$O$_7$ is a free carrier induced `metallic ferroelectric'. Like in undoped A$_3$B$_2$O$_7$ compounds, biaxial strain also modifies the stability range of the polar phase of Ba$_2$SrSn$_2$O$_7$.  

Experimental verification of this prediction is possible. Ba$_3$Sn$_2$O$_7$ is stable in bulk \cite{Hinatsu1998}, and thin films of both Sr and Ba stannate perovskites were grown by multiple groups \cite{Prakash2015, Paik2017, Wang2018}. 
%
%
If the free charge is constrained in the top $\sim$10~\AA~of a Ba$_2$SrSn$_2$O$_7$ film, the charge density needed to stabilize the polar phase is $\sim 5 \cdot 10^{13}$~cm$^{-2}$. Dielectric based gating allow densities of $\sim10^{13}$~cm$^{-2}$ \cite{Goldman2014}, and it is possible to obtain densities exceeding $\sim10^{14}$~cm$^{-2}$ via ionic liquid gating \cite{Yuan2009, Bisri2017, Leighton2019}. 
Thus, it is possible to induce in-plane structural polarization electrolyte gating. The polarization can be observed by second harmonic generation as was done in LiOsO$_3$ \cite{Padmanabhan2018}. 

\textit{Other A$_3$B$_2$O$_7$ compounds --- } 
This mechanism is very general, and it could be expected to be applicable to many other HIF oxides. However, our calculations on Ca$_3$Ti$_2$O$_7$ and Sr$_3$Zr$_2$O$_7$, which we discuss in the supplement \cite{Supplement} indicate that this is not the case. Even though the parent CaTiO$_3$ and SrZrO$_3$ compounds behave very similarly to SrSnO$_3$ under doping, the structural polarization of both Ca$_3$Ti$_2$O$_7$ and Sr$_3$Zr$_2$O$_7$ decrease upon electron doping. The reason is a subtle difference in the nature of polarization in these compounds: While both Ca$_3$Ti$_2$O$_7$ and Sr$_3$Zr$_2$O$_7$ have HIF groundstates, in their $I4/mmm$ reference structure they also display weak polar $\Gamma$ point instabilities \cite{Mulder2016Thesis, Yoshida2018-2}. As a result of this instability, Ca$_3$Ti$_2$O$_7$ has a significant Ti contribution to polarization. This contribution is reduced as electrons are introduced to the system, because free carriers suppress the Ti--O hybridization and harden this soft mode in Ca$_3$Ti$_2$O$_7$ as they do in titanate perovskites CaTiO$_3$ or BaTiO$_3$ \cite{benedek2016ferroelectric, Wang2012, Supplement}. 
Sr$_3$Sn$_2$O$_7$, on the other hand, has no $\Gamma$ instabilities, and has only a negligible SnO$_2$ layer polarization. This suggests the stannate perovskites as a unique group of compounds that can display free carrier enhanced (or induced) hybrid improper ferroelectricity. 

\textit{Summary --- } Using first principles calculations and studying the oxygen octahedral rotations in $Pnma$ perovskites under doping, we showed that the structural polarization in stannate HIFs is not only robust against free carriers, but it is also enhanced. We furthermore predicted a yet-to-be-synthesized compound Ba$_2$SrSn$_2$O$_7$ that undergoes a centrosymmetric to polar transition under electron concentrations that are experimentally achievable by ionic liquid/gel gating. 
Our results show that the improper ferroelectricity driven by steric lattice instabilities can serve as a means to obtain carrier induced ferroelectricity in compounds where those instabilities are strengthened by the free carriers.

\begin{acknowledgments}
We acknowledge helpful suggestions of an anonymous Referee A for the discussion about Ca$_3$Ti$_2$O$_7$. This work was supported primarily by the National Science Foundation through the University of Minnesota MRSEC under Award Number DMR-2011401. We acknowledge the Minnesota Supercomputing Institute (MSI) at the University of Minnesota for providing resources that contributed to the research results reported within this paper.
\end{acknowledgments}

\clearpage
\onecolumngrid

\renewcommand{\theequation}{S\arabic{equation}}
\renewcommand{\thefigure}{S\arabic{figure}}
\renewcommand{\thetable}{S\arabic{figure}}
\renewcommand{\thesection}{S\arabic{figure}}
\renewcommand{\bibnumfmt}[1]{[S#1]}

\setcounter{figure}{1}
\setcounter{table}{1}
\setcounter{section}{1}


\section{Supplementary information}

		\subsection{Methods}

		The Density functional theory calculations are performed using the projector augmented wave approach \cite{blochl1994projector} as implemented in the Vienna Ab-initio Simulation Package (VASP) \cite{VASP1,VASP2}, and using the PBEsol generalized gradient approximation\cite{PBEsol} and a 500~eV plane wave cut-off.
		A $\sqrt{2}\times\sqrt{2}\times2$ supercell, which can capture the $a^-a^-c^+$ octahedral rotation pattern (No.62, $Pnma$), was used for ABO$_3$ perovskites. For A$_3$B$_2$O$_7$ Ruddlesden-Popper structures, a $\sqrt{2}\times\sqrt{2}\times2$ supercell was used. A $6\times6\times4$ k-grid was used for the perovskite supercells, and a $6\times6\times2$ k-grid for the Ruddlesden-Popper supercells was used for all calculations except the DOS calculations which required a finer grid. For structural relaxations, the residual force tolerance was set to 1~meV/\AA. Different space groups are also considered during this process, detailed information can be found in the following sections. In all compounds considered, the octahedral rotation patterns and space groups consistent with previous reports were found to be the lowest energy ones. 
		
		The carrier doping without substituting or adding atoms is simulated by increasing the number of electrons in the DFT calculations. In this approach, in order to ensure the charge neutrality in presence of the free carriers, a uniform background with opposite charge is also added. Since the absolute energy between different doping levels are not compared and only the relative energy with same doping level is important, no energy corrections were made to compensate for this artifact. Most of the trends reported consider ionic relaxations at fixed unit cell volume and shape, and the trends are qualitatively similar when cell volume is relaxed as well.
		
		The Landau free energy expansion was built using the irreducible representations (irreps) of the cubic Pm$\bar{3}$m structure. Irreps that are relevant to the Pm$\bar{3}$m--Pnma phase transition were determined using the ISODISTORT tool of the Isotropy Software Suite \cite{isotropy2007}. The irreps being considered first are $M_3^+(0,0,a), R_5^-(a,a,0)$ and $X_5^-(a,a;0,0;0,0)$. The free energy upto 4th order was considered, and a mesh of finite displacements of these irreps with 10 steps on each direction was used for DFT calculations, which totals up to $10\times10\times10=1000$ data points for one fitting. The parameters of the free energy were fit to the DFT energies on this grid using the standard fitting algorithms as implemented in Numpy \cite{2020SciPy-NMeth}. There are also other irreps including $M_2^+$ and $R_4^-$ present during the phase transition from $Pm\bar{3}m$ to $Pnma$, which are also considered in our calculations by fixing the ratio of $\frac{|M_2^+|}{|M_3^+|}$ and $\frac{|R_4^-|}{|R_5^-|}$. 
		
		Polarization is not a well defined quantity in metals. The amplitude of the polar structural mode is often used as an alternative to the polarization in metallic solids, but this approach does not provide any information on the ionic charges at all. An alternative is to calculate the polarization of the fully filled bands only, or if there are relatively flat bands, to use a generalized Berry phase formalism that takes into account different numbers of bands at each k-point \cite{Filippetti2016}. We use a simplified approach where the polarization is calculated by the product of nominal charges and polar displacements:
		\begin{equation}
			\vec{P} = \sum_i  \vec{u}_i^{\Gamma} Z_i
		\end{equation} 
		where the $\vec{u}_i^{\Gamma} $ represents the displacement vectors of atom $i$ and $Z_i$ is the nominal ionic charge of atom $i$. Using the Born effective charges of the undoped compounds instead of the nominal ionic charges would not lead to a qualitative difference. The nominal, as well as the in-plane Born effective charges in undoped Sr$_3$Sn$_2$O$_7$ are shown in Table.~\ref{tab:charges}. An advantage of this approach, as opposed to just using the polar mode amplitudes without multiplying with charges, is that the acoustic mode (which can be considered as nothing but an origin shift) is taken off by default since the sums of the charges is zero. 

		In order to determine the ground state of the Ruddlesden-Popper structures, several candidate structures with different octahedral rotation patterns are considered. (See Table \ref{tab:trilinear}.) The only structure with a polar point group is $Ama2_1$, and hence we only refer a material as hybrid improper ferroelectric only when the $Ama2_1$ phase has lowest energy. 

                \begin{table}[!]
                        \begin{tabular}{|c|c|c|c|}
                                \hline Atom & Wyckoff position & Nominal charge & Born effective charge (xx-direction) \\ \hline
                                Sr & b & +2 & +2.29 \\
                                Sr & e & +2 & +2.13 \\
                                Sn & e & +4 & +4.03 \\
                                O & a & -2 & -1.50 \\
                                O & e & -2 & -2.99\\
                                O & g & -2 & -1.61  \\ \hline
                        \end{tabular}
                \caption{The nominal and Born effective charges in Sr$_3$Sn$_2$O$_7$.}
                \label{tab:charges}
                \end{table}

\subsection{Other perovskites}
		\begin{figure}[!]
			\includegraphics[width=0.9\textwidth]{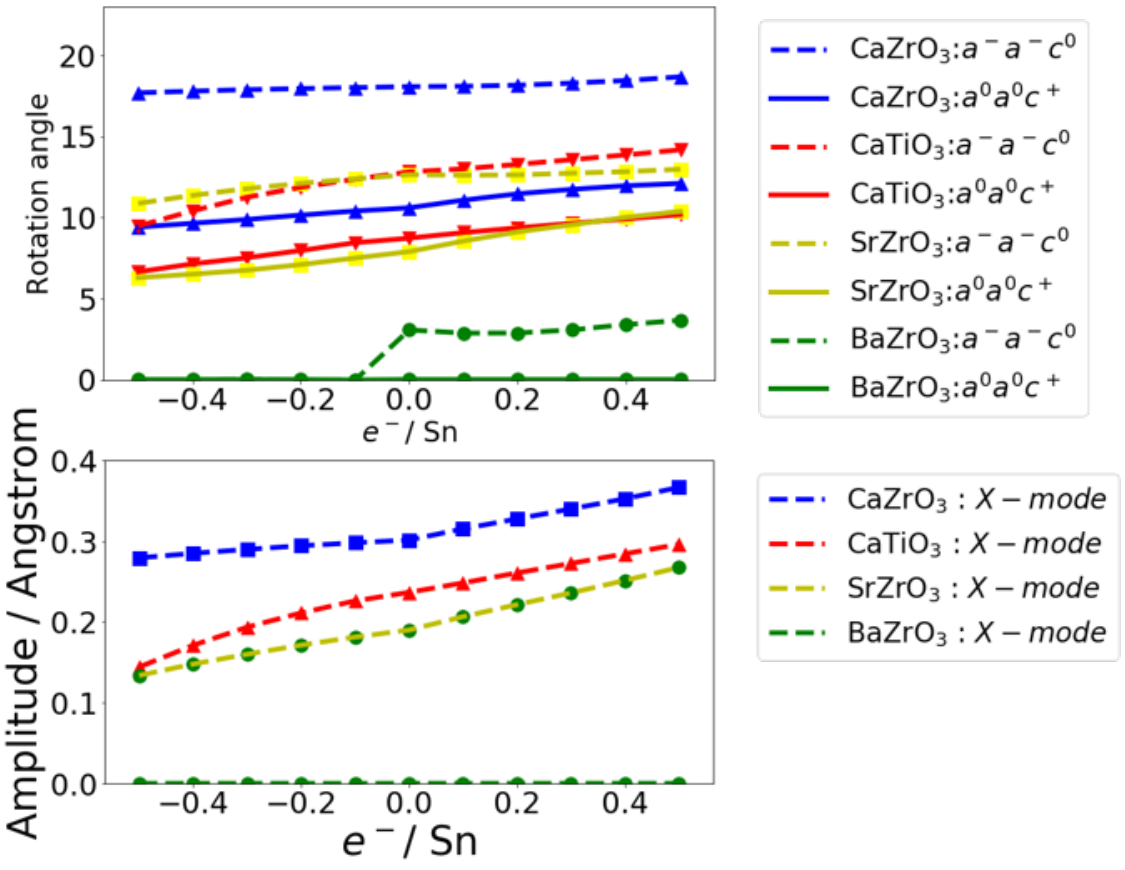}
			\caption{(a) The rotation angles of octahedral rotation modes and (b) the anti-polar mode amplitude increase with increasing electron doping in Zr- and Ti- based perovskites.}
			\label{fig:other_perovskites}
		\end{figure}
		In Fig.\ref{fig:other_perovskites}, the oxygen octahedral rotation angle and anti-polar $X_5^-$ mode amplitudes of Zr- and Ti- based perovskites are shown as a function of carrier doping. Compared with Sn-based perovskites, the distortion modes in these transition metal based perovskites are generally less sensitive to doping. This is likely because the transition metal Zr and Ti ions have unfilled d-orbitals, which form the bottom of the conduction band, and get filled first when electrons are introduced to the system. The lower lying d orbitals are of t2g character, and hence have nodes in the directions of the oxygen ions. In the Sn based perovskites, the bottom of the conduction band is formed by the Sn 5s orbitals, which are larger than the d orbitals, and do not have an nodes. 		
		
		\subsection{Possible structures for RP-phase perovskites}

		There are multiple possible structures for a Ruddlesden-Popper layered perovskites with $t<1$. A systematic group theory analysis had been performed before \cite{Nowadnick2016, Li2020s}. Table.~\ref{tab:trilinear} shows the list of possible space groups for RP-phase perovskites, which includes the parent phase ($I4/mmm$), the polar phase ($Ama2_1$) and other non-polar or anti-polar phases. 
			\begin{table}[!]
			\centering
			\begin{tabular}{|c|c|c|c|}
				\hline
				Irrep 1       & Irrep 2       & Polarization 		& Space  group                                    \\ 
				\hline
				$X_1^- (a,0)$ &               &      Non-polar         & $Aeaa$ (\#68)                                   \\
				$X_2^+ (a,0)$ &               &       Non-polar         & $Aeam$ (\#64)                                     \\
				$X_3^- (a,0)$ &               &        Non-polar        & $Amam$ (\#63)                                       \\
				$X_3^- (a,a)$ &  &  Non-polar        & $P4_2/mnm$ (\#136) 			\\
				$X_1^- (a,0)$ & $X_3^- (0,b)$ & Non-polar        & $Pnab$ (\#60)                                      \\
				
				$X_1^- (a,0)$ & $X_3^- (b,0)$ & Non-polar        & $C/2c$ (\#15)                                    \\
				$X_1^- (a,a)$ & $X_3^- (b,b)$ & Non-polar        & $C/2m$ (\#12)    \\
				
				$X_2^+ (a,0)$ & $X_3^- (b,0)$ & Polar  & $A2_1am$ (\#36)    \\
				$X_2^+ (0,a)$ & $X_3^- (b,0)$ & Anti-polar        & $Pnam$ (\#62)                                     \\  
				
				$X_2^+ (a,a)$ & $X_3^- (b,b)$ & Non-polar        & $C2mm$ (\#38) \\
				\hline
			\end{tabular}
				\caption {List of structures that can be obtained by combining the unstable $X$ modes: $X_1^-$ modes are the out-of-phase rotation along the c-axis, $X_2^+$ modes are in-phase rotation along the c-axis, $X_3^-$ modes are out-of-phase rotation along the a- or b- axis. There can be trilinear-coupled phonon modes at $\Gamma$ or $M$ points if two $X$-modes are present. These $\Gamma$ or $M$ modes are mostly non-polar mode except $\Gamma_5^-$, a complete list of these coupled modes can be found in an earlier work\cite{Li2020s}. While we performed DFT calculations for the energies of all of these phases, only the ones that are close to the lowest energy are shown in the plots. }
			\label{tab:trilinear} 		
		\end{table}
	
		In our calculations of phase diagram, all the phases listed in Table.~\ref{tab:trilinear} have been considered, but only the lowest energy structures are shown. For Sr$_3$Sn$_2$O$_7$, the lowest energy structure is always the polar one when electron doped. However, the biaxial strain could induce a polar-nonpolar phase transition.  For SrBa$_2$Sn$_2$O$_7$, the free electron charge carrier can induce a polar-nonpolar phase transition, which has been discussed in the main text.  
		
		\subsection{Phase diagram of Sr$_3$Sn$_2$O$_7$ under hole doping.}
		\begin{figure}[!]
			\includegraphics[width=0.9\textwidth]{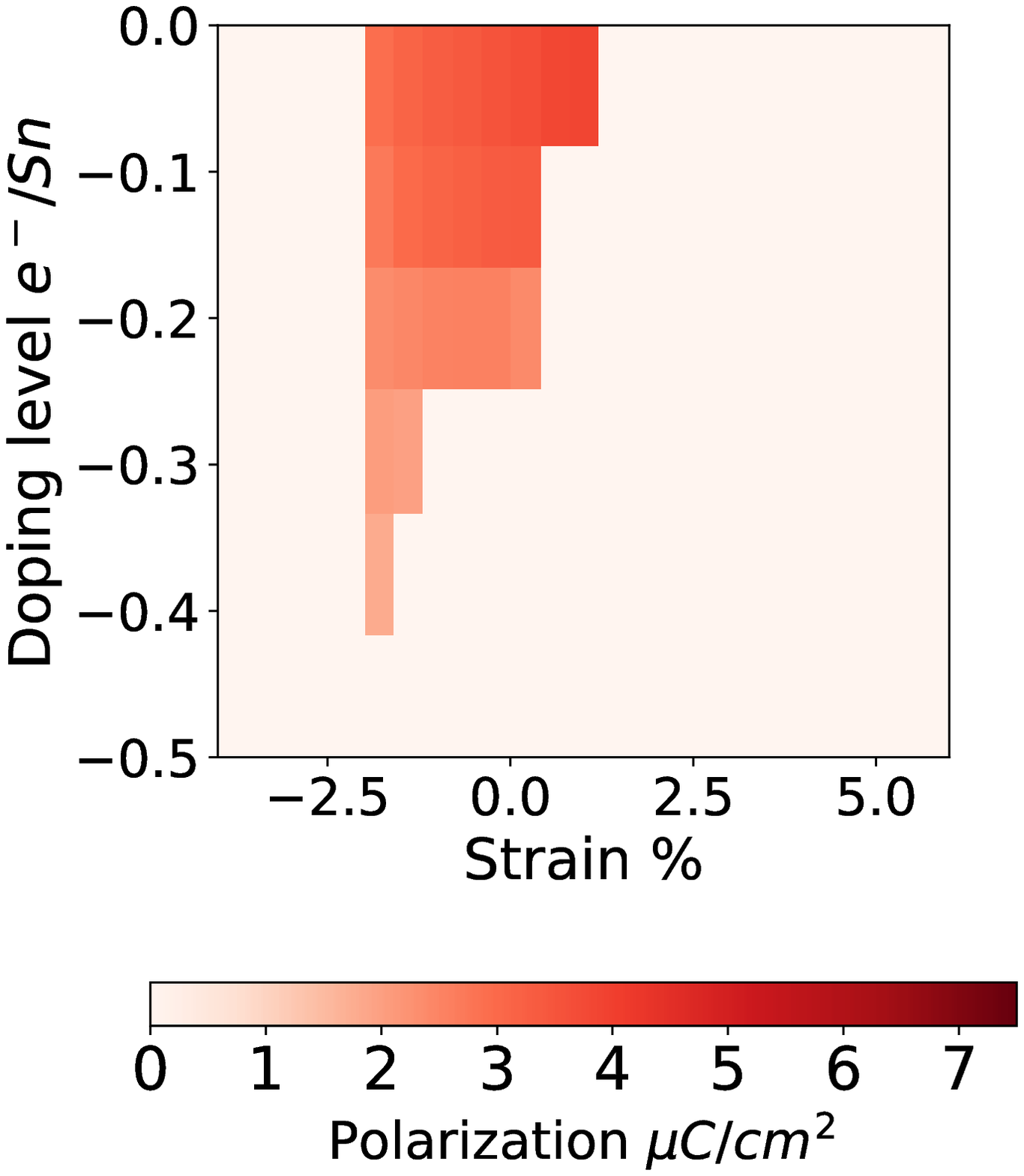}
			\caption{The structural polarization of Sr$_3$Sn$_2$O$_7$ when hole doping.}
			\label{fig:phase_negative}
		\end{figure}
		Fig.~\ref{fig:phase_negative} shows the phase diagram of  Sr$_3$Sn$_2$O$_7$ when hole doped. Unlike BaTiO$_3$\cite{Wang2012s, Hickox-Young2020s} where the polarization is screened with either hole or electron charge carriers, only the hole doping is detrimental to the hybrid improper ferroelectricity in  Sr$_3$Sn$_2$O$_7$. Holes are introduced to the O-2p orbitals, which decreases the effective ionic radii of the oxygen ions. This decreases the mismatch between the A-site ions and BO$_6$ octahedra, i.e. tolerance factor. (This is because the first derivative of the tolerance factor with respect to the oxygen radius $R_O$ is $\partial \tau/\partial R_O \propto (R_B-R_A) < 0$.) Hence the rotations of the BO$_6$ octahedra are decreased.

		\subsection{Strain-induced phase transition}
		\begin{figure}[!]
			\includegraphics[width=0.9\textwidth]{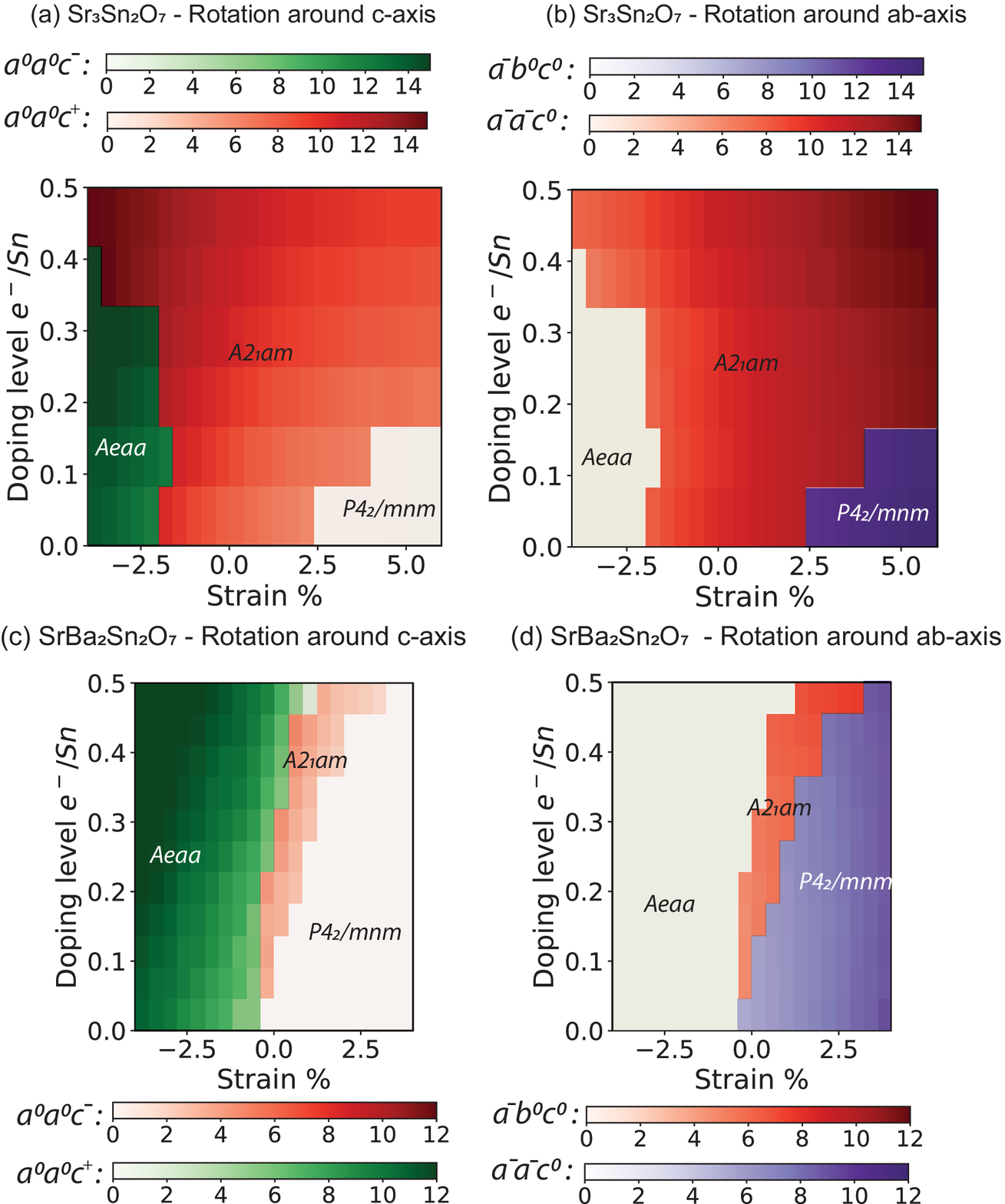}
			\caption{The rotation angles of oxygen octahedra in (a-b)Sr$_3$Sn$_2$O$_7$ and (c-d) SrBa$_2$Sn$_2$O$_7$. (a)(c) are rotations around the c-axis while (b)(d) are rotations around the a ([001])  or ab ([110]) axis. The red region is the a polar phase, which has two rotation modes: $X_2^+(1,0)$($a^0a^0c^+$) and $X_3^-(1,0)$($a^-a^-c^0$). The green region is a non-polar phase, which has only one rotation mode: $X_1^-(1,0)$($a^0a^0c^-$). The purple region is also a non-polar phase, which has one rotation mode: $X_3^-(1,1)$($a^-b^0c^0$).}
			\label{fig:rotation_phase}
		\end{figure}
		There are two non-polar phases in the epitaxial-strain phase diagrams of Sr$_3$Sn$_2$O$_7$ and SrBa$_2$Sn$_2$O$_7$: $Aeaa$ in the compressive strain regime and $P4_2/mnm$ in the tensile strain part. (Here we focus on the cation ordering pattern that does not break the $I4/mmm$ symmetry in SrBa$_2$Sn$_2$O$_7$.) These phases' ranges of stability are shown in the main text Fig.3(c) and Fig.4(b). As shown in Table.~\ref{tab:trilinear}, the polar phase $A2_1am$ has two rotation modes: the $X_2^+(1,0)$ is an in-phase rotation around the c-axis ($a^0a^0c^+$) and $X_3^-(1,0)$ is an out-of-phase rotation around the inplane a- and b-axes  ($a^-a^-c^0$). (Throughout this manuscript, we use the axes of the tetragonal conventional cell of the reference high symmetry structure $I4/mmm$. Under biaxial strain, the double-tilt system transforms into two single tilt systems: $Aeaa$ in the compressive strain side and $P4_2/mnm$ in the tensile strain side. In $Aeaa$, the rotations around the in-plane axes disappear and the rotation around c-axis become out-of-phase, which is $X_1^-(1,0)$ ($a^0a^0c^-$). Similarly, in $P4_2/mnm$, the rotation around c-axis disappear while the rotations around [110] axis now become rotation around [100] axis, which is the $X_3^-(1,1)$ mode ($a^-b^0c^0$) \cite{Li2020s}. 
		
		Possibly the simplest phenomenological explanation of the transition to a single tilt system under compressive strain relies on the observation that in most oxide perovskites the effect of compressive strain is to strengthen the rotation around c-axis, while the rotation angles around a- and b- axes decrease, as shown in the Fig.~\ref{fig:rotation_phase}. Consider the free energy expansion of these two rotation modes: 
		\begin{equation}
			\mathbf{F} = \alpha_R R^2 + \alpha_T T^2 + \beta R^2 T^2
		\end{equation} 
		where $R$ is the rotation amplitude around the c-axis and $T$ is the rotation (tilt) amplitude around the in-plane axes. Both $\alpha_R$ and $\alpha_T$ are less than zero because they are unstable according to the phonon calculations, but $\beta$ is positive since the rotations compete with each other. Under compressive strain, when $R$ becomes large enough so that 
\begin{equation} 
R^2>\frac{|\alpha_T|}{\beta}  
\label{equ:border}
\end{equation}
T would be supressed \cite{Li2020s}. 
		
This simple model also explains why the nonpolar phase that emerges under compressive strain is only stable at lower electron doping, albeit only qualitatively. Under electron doping, the second-order coefficients get significantly enhanced (more than 100\% when $0.5e^-$ per Sn-site) as shown in the Fig.~\ref{fig:landau_327}, and the ratio $\alpha_T/\alpha_R$ increases. This makes the in-plane tilts (T) more stable against the biquadratic coupling $\beta$, since right hand side of Eq.~\ref{equ:border} increases under doping faster than the right left hand side. This is why the stable region of the double tilting modes keep increasing with electron doping, as observed in the main text Fig.3(c) and Fig.4(b).

\subsection{Unstable Phonons of Sr$_3$Sn$_2$O$_7$ Under Doping}

\begin{figure}[!]
\includegraphics[width=0.7\textwidth]{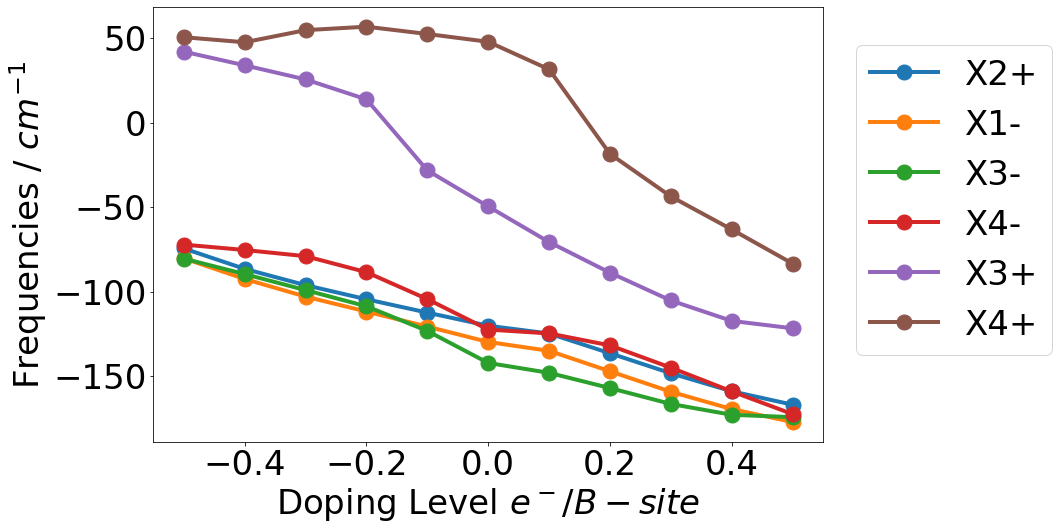}
\caption{Frequencies of the unstable X point modes of Sr$_3$Sn$_2$O$_7$ under carrier doping.} 
\label{fig:Sr3Sn2O7PhononsDoping}
\end{figure}

In Fig. \ref{fig:Sr3Sn2O7PhononsDoping}, we display the frequencies of the unstable phonon modes at the X point of Sr$_3$Sn$_2$O$_7$, and their evolution with the introduction of free carriers. Following earlier work, we have considered a range of structures that can be obtained by combinations of $X_3^-$, $X_1^-$, and $X_2^+$ modes. The $X_4^-$, which is unstable is ignored because it is always weaker than the $X_3^-$ instability, which has a similar but more favorable distortion pattern. (It has been discussed, for example, for Sr$_3$Zr$_2$O$_7$ and not lead to a phase that competes with the groundstate \cite{Yoshida2018-2s}.) Similarly, $X_3^+$ (which is much less unstable) and also $X_4^+$ (which becomes unstable only under very large values of electron doping) are not taken into account in the symmetry analysis.

\subsection{Other RP-phase perovskites}
		\begin{figure}[!]
			\includegraphics[width=0.6\textwidth]{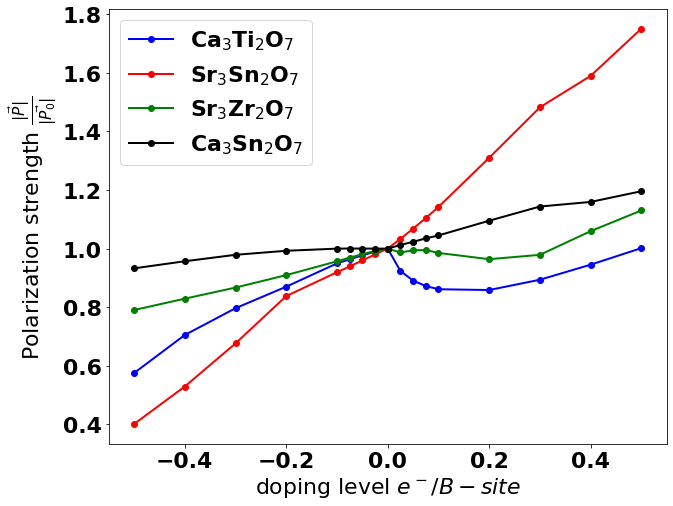}
			\caption{Charge carrier doping will change the structural properties. The net polarization of four different RP-phase perovskites under doping.}
			\label{fig:other_RP}
		\end{figure}
	
Unlike the tin-based perovskites like Sr$_3$Sn$_2$O$_7$, where an immediate upward trend of polarization is seen when free electrons are introduced to the system, the polarizations of Zr- and Ti- based RP-phase perovskites initially decrease under electron doping (Fig.~\ref{fig:other_RP}). This is despite the fact that all of these elements' ABO$_3$ perovskites display rotation trends similar to each other, and that at higher doping concentrations an upward trend in polarization is observed in all of them. In this section, we focus on Ca$_3$Ti$_2$O$_7$ to explain the reasons underlying the seemingly unique behavior of stannates.  
		
\begin{table}[]
	\begin{tabular}{|p{0.2\textwidth}p{0.15\textwidth}p{0.15\textwidth}|}
		\hline
		& Sr$_3$Sn$_2$O$_7$ & Ca$_3$Ti$_2$O$_7$ \\ \hline
		$\Gamma_5^-$ mode-A & 93.91 $cm^{-1}$       & \textbf{-61.71   $cm^{-1}$ }    \\ 
		$\Gamma_5^-$ mode-B & 69.87 $cm^{-1}$       & 75.05  $cm^{-1}$        \\  \hline
	\end{tabular}
\caption{The phonon frequencies of polar displacement modes shown in Fig.~\ref{fig:phonon_SSO_CTO} in undoped Sr$_3$Sn$_2$O$_7$ and Ca$_3$Ti$_2$O$_7$.}
\label{tab:phonon_SSO_CTO}
\end{table}

\begin{figure}[!]
	\includegraphics[width=0.9\textwidth]{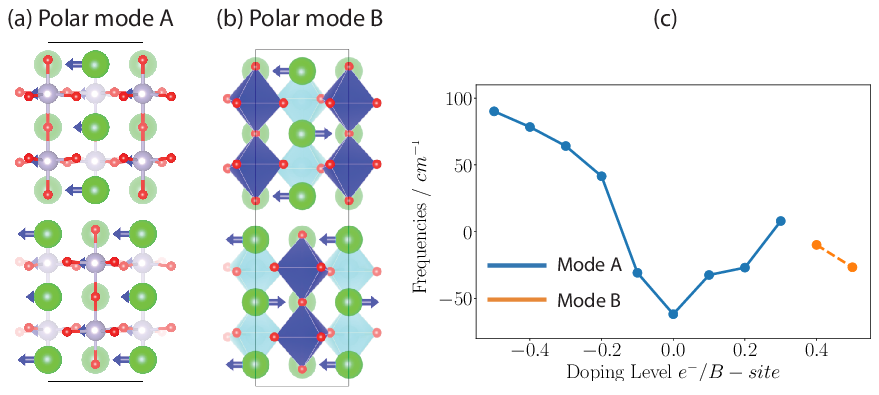}
	\caption{In the undoped Ca$_3$Ti$_2$O$_7$, there are two origins of the polarization: (a) The proper ferroelectric mode and (b) hybrid improper ferroelectric mode that coupled with rotation modes. (c) The frequency of the polar mode (schematic on the left) in Ca$_3$Ti$_2$O$_7$ change rapidly under charge carriers doping. Note that especially for large electron doping the characters of these modes mix, and we classified them as mode-A or mode-B according to the direction of the inner vs. vacuum AO layers' displacements being parallel or antiparallel.}
	\label{fig:phonon_SSO_CTO}
\end{figure}

Both CaTiO$_3$ and SrSnO$_3$ have the same $Pnma$ structure with $a^-a^-c^+$ octahedral rotation pattern, but there is a crucial difference between these two compounds: While cubic SrSnO$_3$ has no $\Gamma$ point instability, CaTiO$_3$ has a polar unstable mode in its cubic phase, which gets suppressed through the biquadratic interactions with the octahedral rotation modes \cite{Benedek2013}. This absence of a polar instability in SrSnO$_3$ can be explained by 1) the absence of significant Sn-O hybridization due to the larger bandgap, and 2) the larger tolerance factor of SrSnO$_3$ compared to CaTiO$_3$, which makes the A-site better coordinated and hence lowers the tendency towards A-site ferroelectricity.
The A$_3$B$_2$O$_7$ Ruddlesden-Poppers also behave similarly: Sr$_3$Sn$_2$O$_7$ in the $I4/mmm$ reference structure has no $\Gamma$ instability, whereas Ca$_3$Ti$_2$O$_7$ has a polar instability as shown in Table~\ref{tab:phonon_SSO_CTO} \cite{Mulder2016Thesiss}. (Sr$_3$Zr$_2$O$_7$, which behaves similar to Ca$_3$Ti$_2$O$_7$ also has a $\Gamma$ instability \cite{Yoshida2018-2s}.)
The character of the unstable mode in undoped Ca$_3$Ti$_2$O$_7$, which we refer to as `mode-A', is not surprising, and it is similar to the polar mode in CaTiO$_3$: It consists of significant and parallel Ca displacements, that are accompanied with Ti displacements (Fig. \ref{fig:phonon_SSO_CTO}). The next lowest frequency polar mode in Ca$_3$Ti$_2$O$_7$, which we refer to as `mode-B' has a different character: It has significant anti-parallel Ca displacements, with minimal Ti displacement. 

\begin{figure}[!]
			\includegraphics[width=0.9\textwidth]{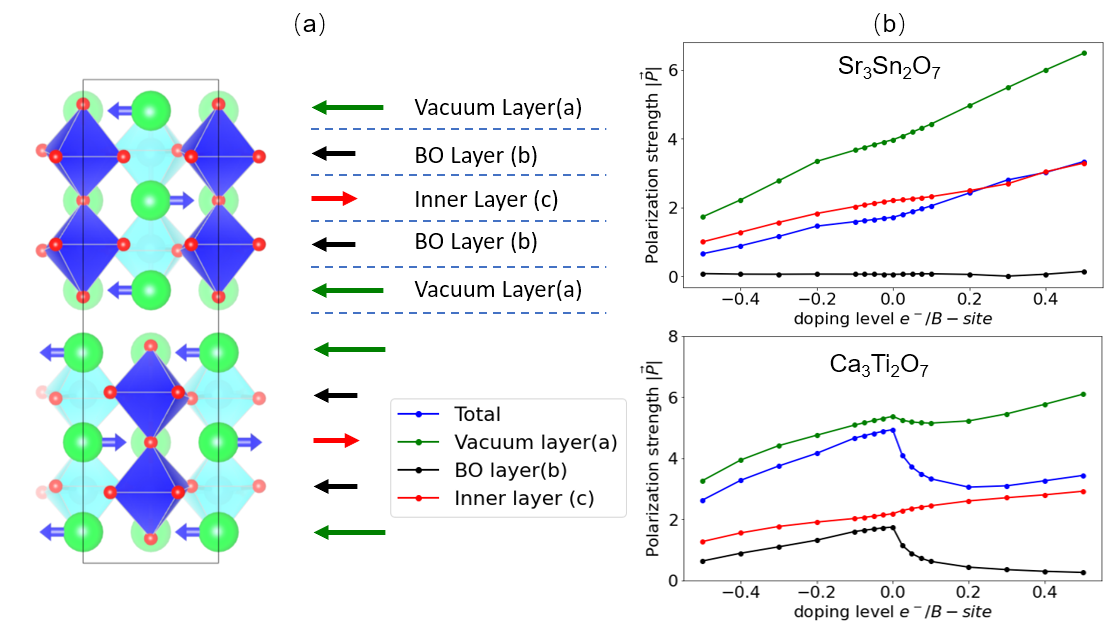}
			\caption{(a) The polarization direction of each layer alternates in HIF A$_3$B$_2$O$_7$ perovskites. The AO rock-salt like layer is noted as vacuum layer, while the other AO layer in the middle of perovskite layer is called inner layer. (b) The polarization strength by layer for Sr$_3$Sn$_2$O$_7$ (top) and Ca$_3$Ti$_2$O$_7$ decomposed by layers.}
			\label{fig:SSO_CTO}
\end{figure}

As a result of these anti-parallel A-site displacements, mode-B is expected to couple more strongly with the octahedral rotations at the trilinear order. As a result, the polar ground state structure of Ca$_3$Ti$_2$O$_7$ has significant contributions from both of these polar modes, which can be seen from the large contribution from both TiO$_2$ and CaO layers to the total polarization (Fig.~\ref{fig:SSO_CTO}). In Sr$_3$Sn$_2$O$_7$, on the other hand, mode-B is softer than mode-A, and as a result, there is almost no contribution from the SnO layers to the polarization as shown in Fig.~\ref{fig:SSO_CTO}. This difference is responsible of the different trends in polarization of these compounds under doping: While the AO layers' contributions to the polarization in both compounds increase under electron doping, the TiO$_2$ contribution from Mode A is suppressed by the introduced electrons. This is expected, since added electrons to the Ti d orbitals are well known to suppress Ti--O hybridization and harden polar soft modes \cite{benedek2016ferroelectrics}. 
The frequency of the unstable phonon mode in tetragonal Ca$_3$Ti$_2$O$_7$ supports this picture: It is hardened under both electron and hole doping. Under large ($\gtrsim  0.3$) electron concentration the polar mode softens with increasing concentration again. This is due to mode-B softening and mixing with mode-A to make the instability domininantly like mode-B at larger dopings, which explains the upturn in polarization in this doping range.

\subsection{Landau analysis}
		\label{sec:landau}
		\begin{figure}
			\includegraphics[width=\textwidth]{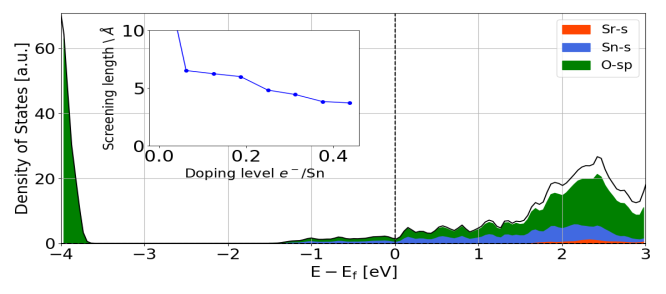}
			\caption{The projected density of states of SrSnO$_3$ when 0.3$e^-$ per Sn atom doped. The inset is the screening length as a function of the doping level. The screening length is calculated using Thomas-Fermi model: $\lambda = \sqrt{\varepsilon /e^2 D(E_f)}$, here the $\varepsilon$ is the dielectric constant of undoped SrSnO$_3$.}
			\label{fig:DOSs}
		\end{figure}
		
		The geometric effects of free electrons can be attributed to two sources: the volume effect and increment of incompatibility (tolerance factor).  When free electrons are introduced to the system, the structure expands, and the majority of electrons move into the oxygen and Sn-5s orbitals and increase their ionic radii. (See the DOS in Fig.~\ref{fig:DOSs}.) This increase can be seen in the derivatives of the tolerance factor with respect to the ionic radii: 
\begin{equation}
t =\frac{R_A+R_O}{\sqrt{2}\left(R_B+R_O\right)}
\end{equation}
\begin{equation}
\frac{dt}{dR_B}=-\frac{R_A+R_O}{\sqrt{2}\left(R_B+R_O\right)^2}
\end{equation}
\begin{equation}
\frac{dt}{dR_O}=\frac{R_B-R_A}{\sqrt{2}\left(R_B+R_O\right)^2}
\end{equation}
An increase in the ionic radius of the B cation necessarily reduces the tolerance factor, and the same applies to the ionic radius of Oxygen as long as $R_B-R_A<0$, which is the case in practically all perovskites. This result implies that in all oxide perovskties where the density of states near the fermi level does not have any contribution from the A-site cation, the effect of electron doping is an effective reduction in the tolerance factor.

		\begin{figure}[!]
			\includegraphics[width=.7\textwidth]{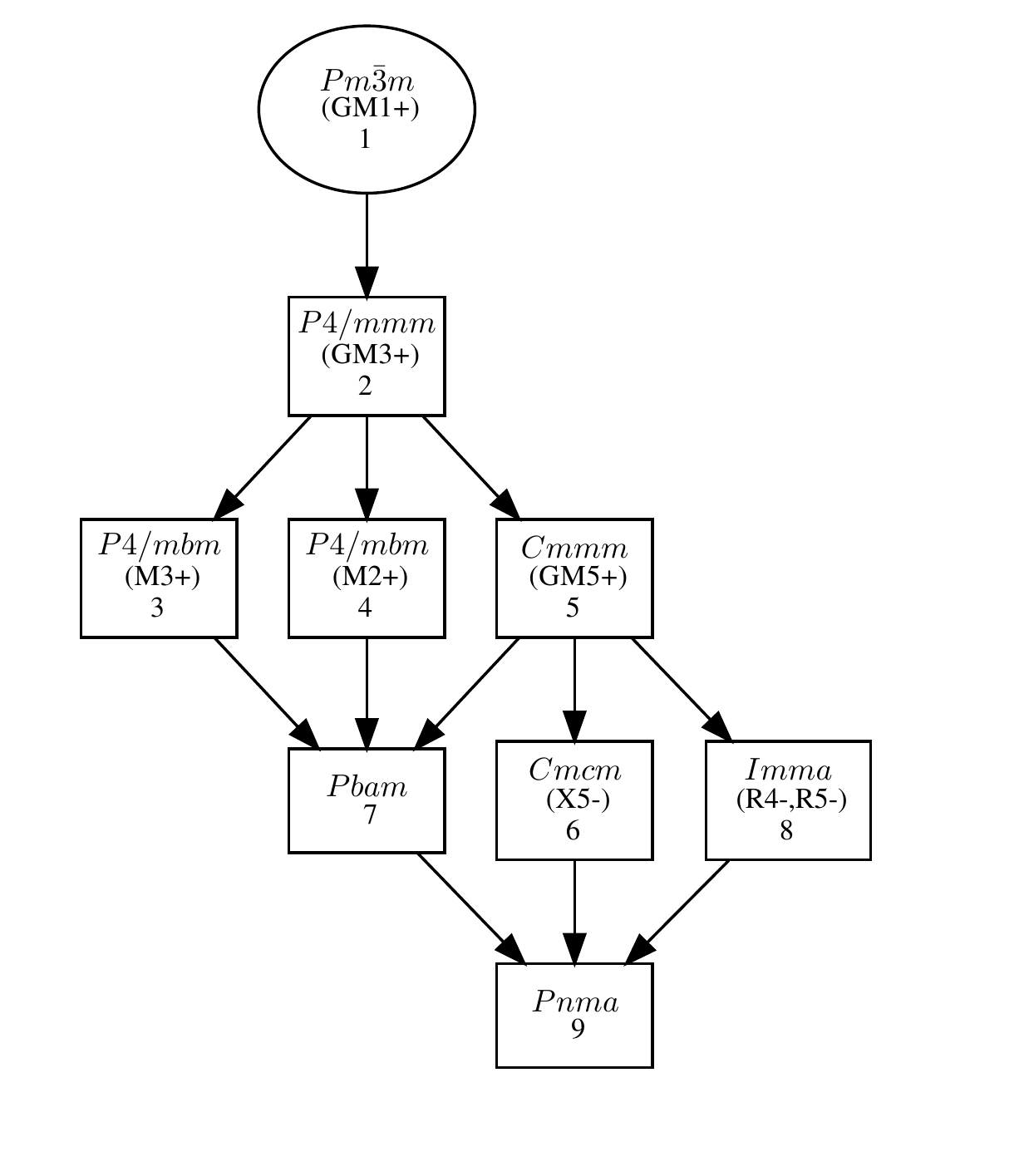}
			\caption{The space groups and distortion modes associated with the phase transition from $Pm\bar{3}m$ to $Pnma$ . This graph is made through Subgroups from Bilbao Crystallographic Server \cite{Ivantchev:ks0038}.}
			\label{fig:phase_transition}
		\end{figure}

		These two effects (volume and tolerance factor change) coexist, but the phonon calculations show that the anti-polar modes get strengthened along with the rotation modes in the Sn based compounds. In order to disentangle the volume expansion effect, we performed DFT calculations in three different configurations: 1) In the first set of calculations, we kept the volume fixed to that of the undoped cubic reference structure. 2) We then repeated the calculations where we relaxed the volume of the cubic cell at each different doping level. 3) For control, we also performed calculations where we used the volumes from step 2, but did not consider any electron doping. (This last step shows the effect of volume expansion only.  
		A phenomenological model is built from Landau theory and energies from DFT calculations. The Landau energy expression that describes the transition from $Pm\bar{3}m$ to $Pnma$ phase is:
		\begin{equation}
		\mathcal{F}=\alpha_R R^2  +\beta_R R^4 +\alpha_M M^2 +\beta_M M^4 +\alpha_X X^2 +\beta_X X^4 +\gamma R\cdot M\cdot X 
		\label{equ:freeenergys}
		\end{equation}
		Here the $R$ represents the amplitude of distortion mode $R_5^-(a, a, 0)$, $M$ represents the amplitude of mode $M_3^+(0,0,a)$, and $X$ represents that of $X_5^-$. These three distortion modes have the greatest amplitudes in the $Pnma$ phase, and the $R$ and $M$ modes are the primary order parameters, but multiple other modes are also present in the $Pnma$ structure as secondary order parameters, as shown in fig.~\ref{fig:phase_transition}.  
		
\begin{figure}[!]
			\centering
			\makebox[\textwidth][c]{
				\centering
				\includegraphics[width=1.\textwidth]{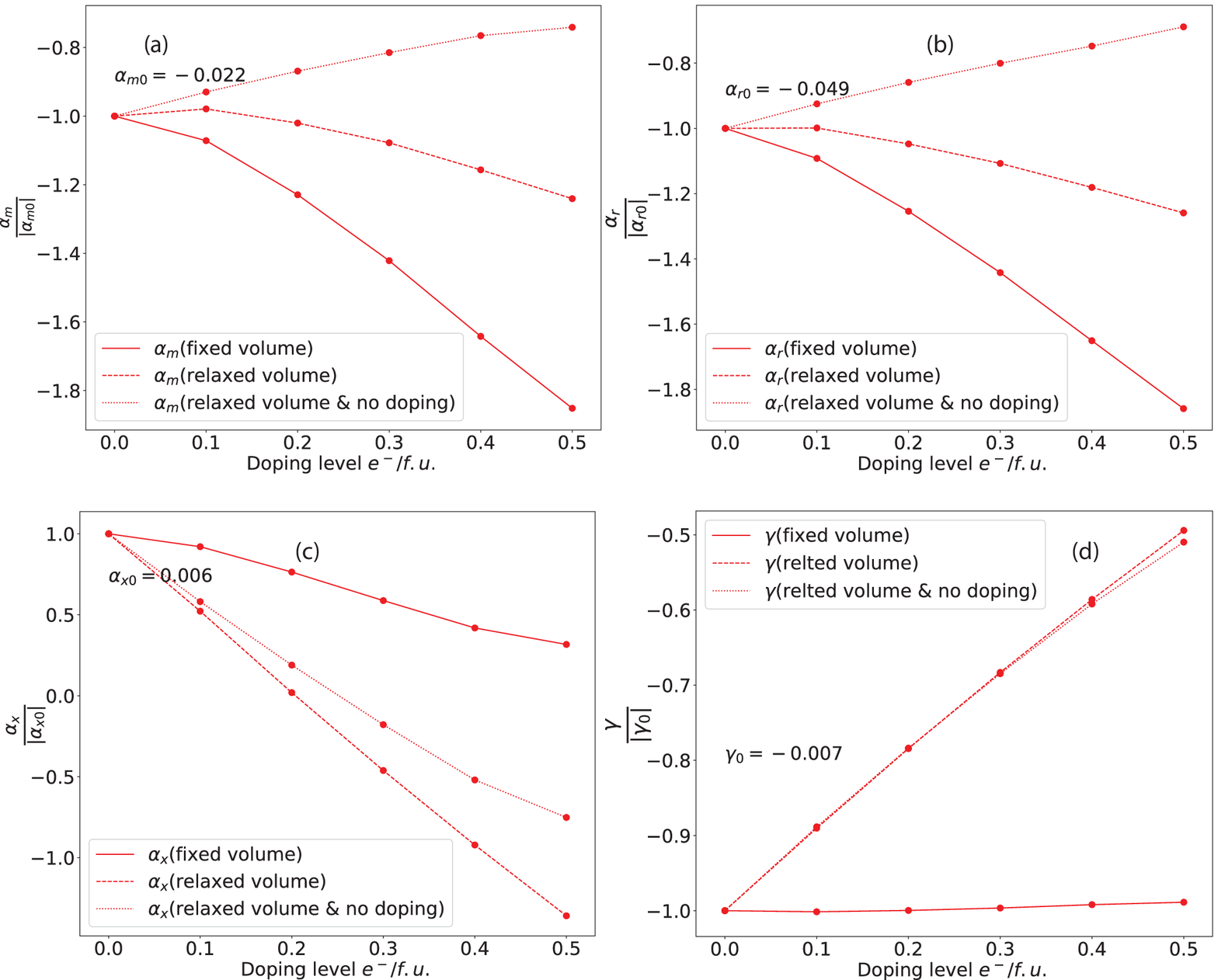}
			}
			\caption{The coefficients of the Landau free energy expansion of SrSnO$_3$ (Eq.~\ref{equ:freeenergys}). When extra electrons are introduced, all rotation modes get softened while the trilinear coupling term remains a positive constant. Three different configurations are shown here, solid: relaxed the atomic position with fixed lattice constant when not doping, coarse dashed: relaxed both atomic positions and lattice constant when doping, fine dashed: relax the atomic positions using the lattice constant when doping, but no free charge carriers are present. Unit of $\alpha$ is $eV/$\AA$^2$, unit of $\gamma$ is $eV/$\AA$^3$.}
			\label{fig:landau}
		\end{figure}

		We calculate the coefficients in the equation.~\ref{equ:freeenergys} by fitting the energy when the structures manually distorted by different amplitudes of normal modes. 10 different amplitudes for distortion modes $R_5^-(a, a, 0), M_3^+(0,0,a), X_5^-$ are applied in this process, thus $10\times10\times10=1000$ energies were used to fit each set of coefficients. Fig.~\ref{fig:landau} shows the change of coefficients as a function of doping level, each data point represents a fitting result from $1000$ different structures. $3\times5\times1000=15000$ different structures were used and calculated to get fig.~\ref{fig:landau}.  
				
		The volume expansion caused by the introduction of free electrons influences the rotation amplitude, but it is not the dominant factor. It is the change in the tolerance factor that dominates the trends. This can be seen from the trends of the (fixed volume) and dashed (relaxed volume) lines in Fig.\ref{fig:landau}a-c. The coefficients of quadratic energetic terms both increase with or without volume changed. The volume change by itself (dotted lines) leads to an opposite trend for $\alpha_M$ and $\alpha_R$. The only exception where relaxing the volume makes a qualitative difference is in the trilinear coupling $\gamma$: This coefficient is almost independent of the free carrier concentration, but reduces rapidly when volume expansion is taken into account.   
		
		Compared with the rotational modes, the anti-polar mode is more sensitive to the volume effect but less sensitive to charge doping. The anti-polar mode have a positive phonon frequency and quadratic energy contribution at the undoped state. The sign of phonon frequency and quadratic energy term changes when the volume expands - this will hugely increase the amplitue of anti-polar mode. The trilinear coupling is also very sensitive to the volume effect despite the sign remaining unchanged. Interestingly, once the volume is fixed, it is almost independent of the doping level.

		In order to estimate the effect of secondary order parameters $M_2^+$ and $R_5^-$ modes, we also performed a separate set of calculations where we considered nonzero amplitudes of these modes as well. Since considering all different values of these modes' amplitudes would make the calculations prohibitive, we fixed the $\frac{|M_3^+|}{|M_2^+|}=2$ and $\frac{|R_4^-|}{|R_5^-|}=0.5$, and then re-fitted the coefficients of the Landau model. These ratios are determined by their value in the undoped-ground state structure. Trends that are qualitatively very similar to those in Fig.~\ref{fig:landau} were found. 
	
	\begin{figure}[!]
		\centering
		\makebox[\textwidth][c]{
			\centering
			\includegraphics[width=1.\textwidth]{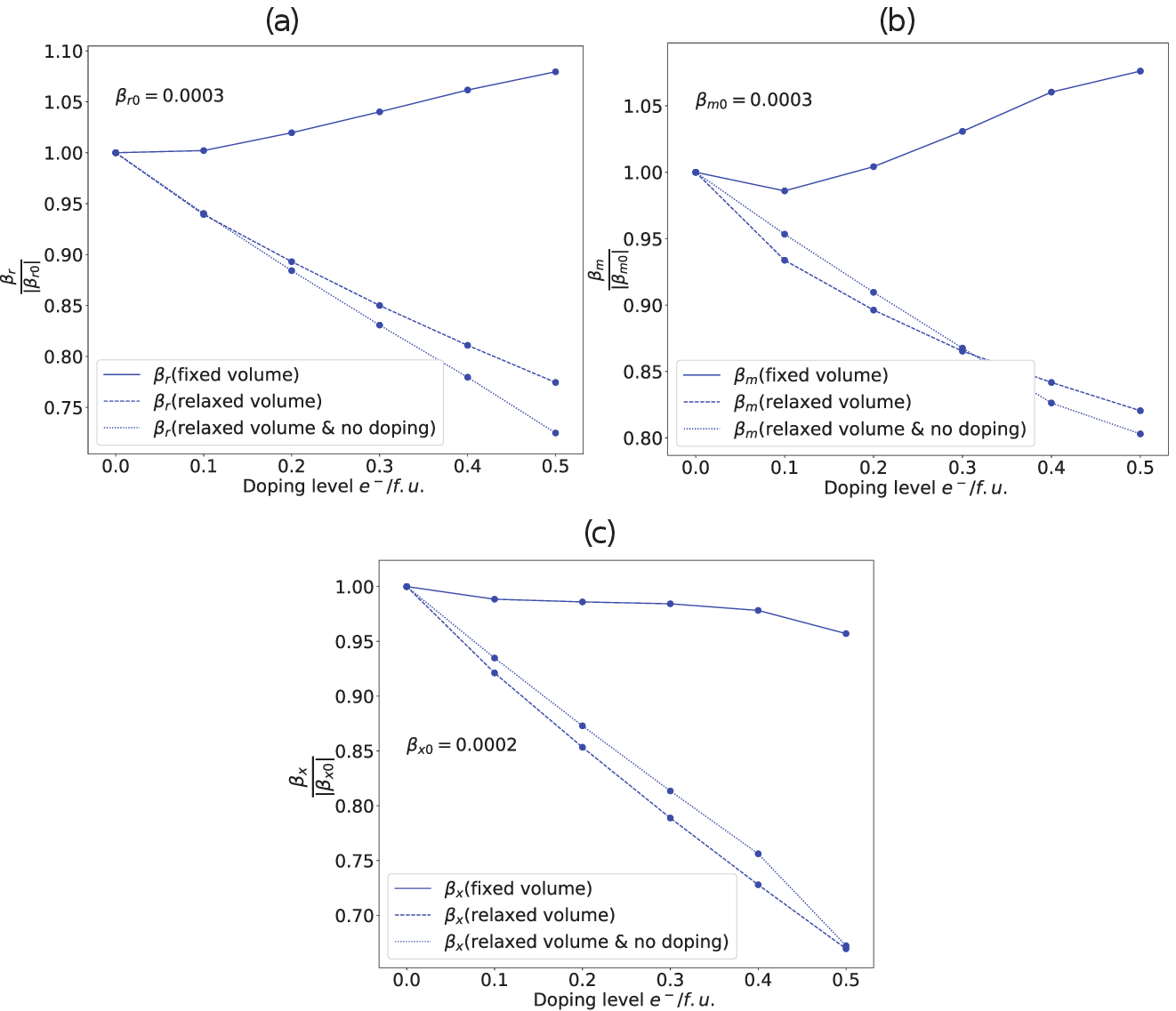}
		}
		\caption{The 4th-order terms' coefficients in landau free energy expansion of SrSnO$_3$.  Unit of $\beta$ is $eV/$\AA$^4$.}
		\label{fig:landau4}
	\end{figure}
		
In Fig.~\ref{fig:landau4}, we show the change of the 4$^{th}$ order terms' coefficients in the free energy expansion. These coefficients barely change (less than $10\%$) when the volume is fixed. Even when the volume is relaxed, the changes in 4th-order terms are less significant compared to the lower order coefficients' effects. 
		
	\begin{figure}
		\centering
		\includegraphics[width=.5\textwidth]{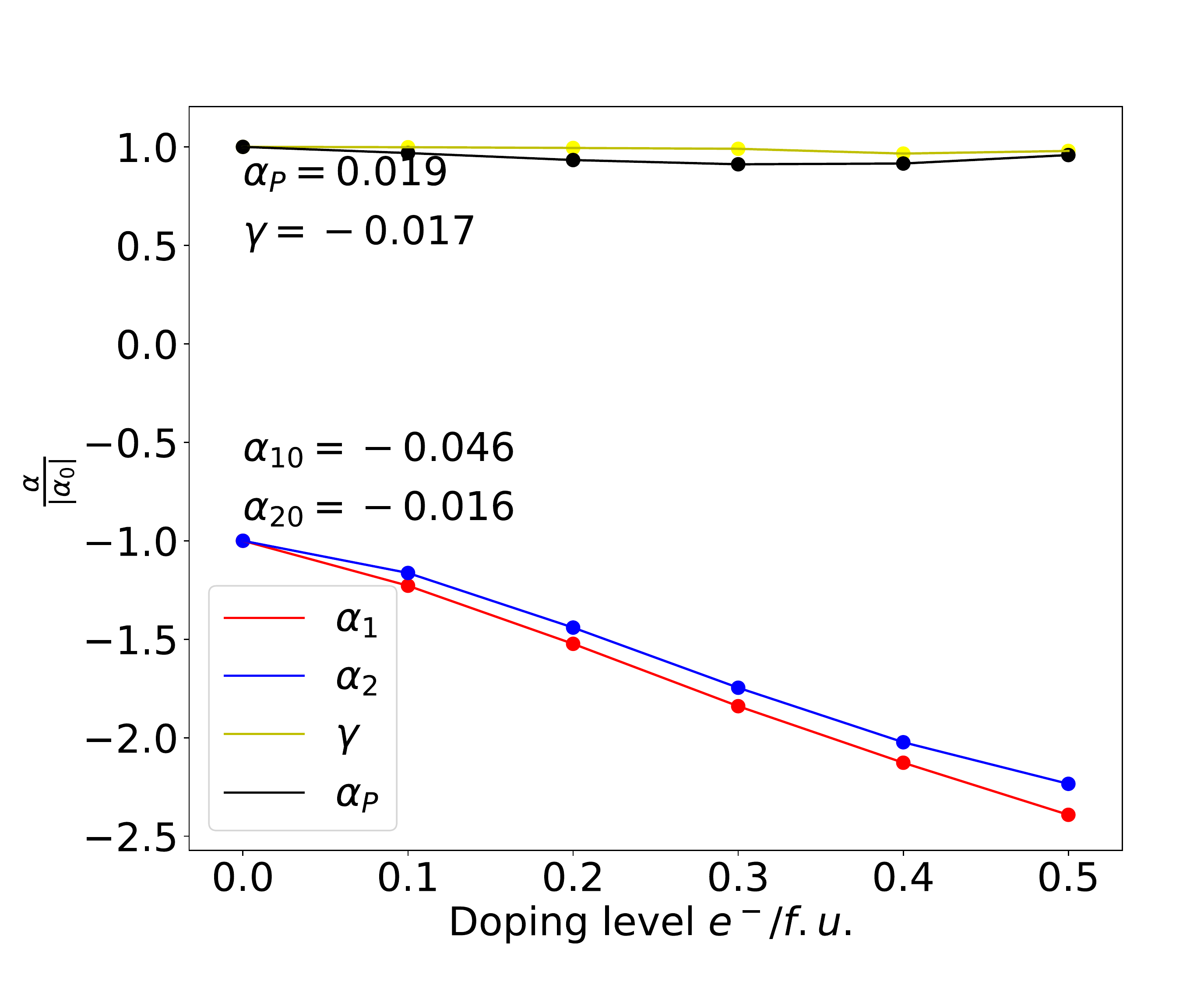}
		\caption{The landau energy expansion coefficients of Sr$_3$Sn$_2$O$_7$, the volume is fixed in this calculation. The landau energy in Sr$_3$Sn$_2$O$_7$ is very similar to that in SrSnO$_3$: $\mathcal{F}=\alpha_{1} Q_1^2 +\alpha_{2} Q_2^2 +\alpha_P P^2 + \gamma Q_1 Q_2 P $, where $Q_1, Q_2, P$ stands for the amplitudes of $X_3^-, X_2^+, \Gamma_5^-$.}
		\label{fig:landau_327}
	\end{figure}

Finally, in Fig.~\ref{fig:landau_327} we show the trends in the coefficients in the free energy expansion of Sr$_3$Sn$_2$O$_7$ that are relavant to the transition from $I4/mmm$ to $Ama2_1$. The trends are similar to those observed in SrSnO$_3$.

\subsection{Volume Effect on Sr$_3$Sn$_2$O$_7$}
		\begin{figure}[!]
		\centering
		\includegraphics[width=.5\textwidth]{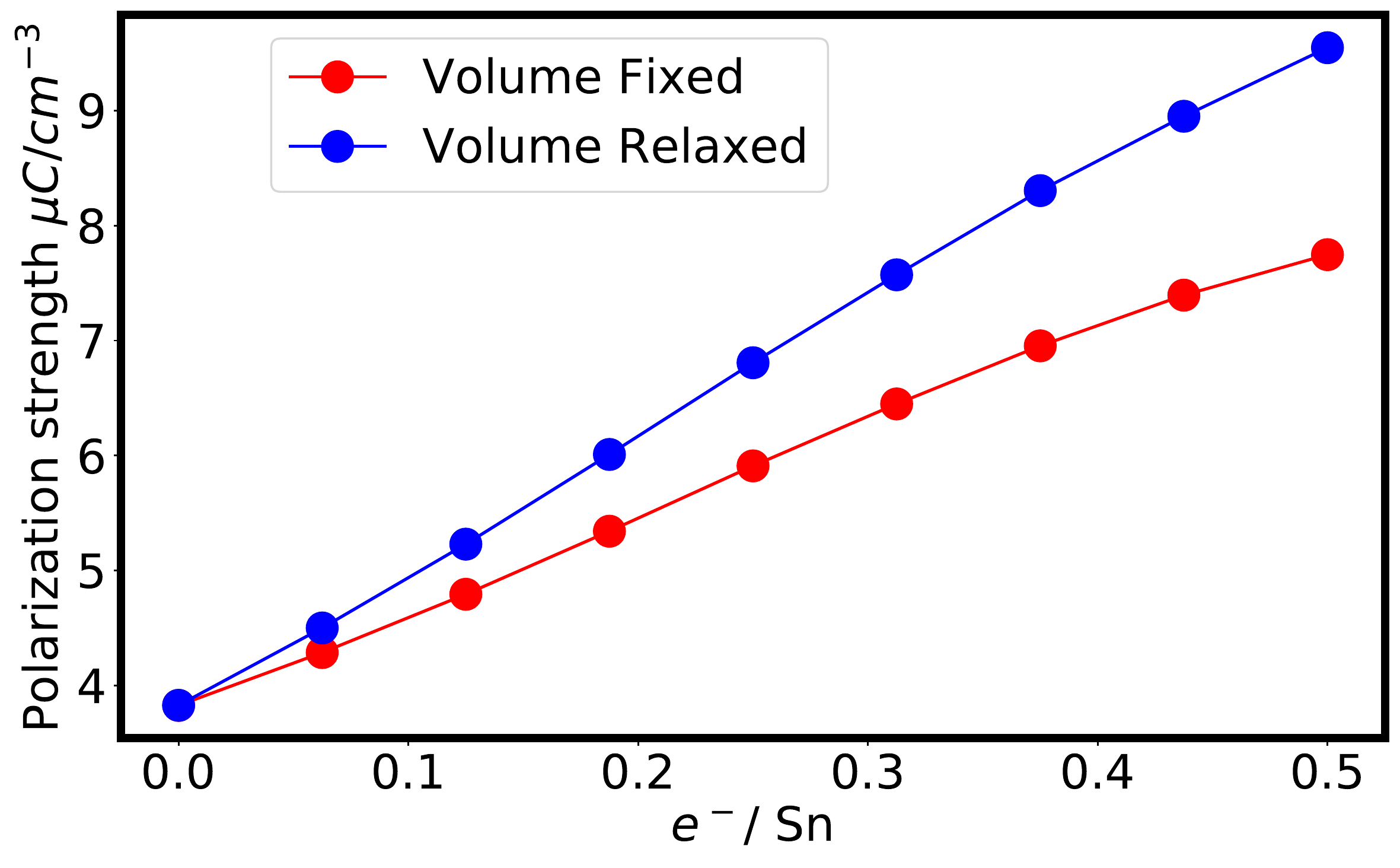}
		\caption{The polarization of Sr${_3}$Sn$_2$O$_7$ compounds as a function of doping level, with and without the volume fixed. It can be seen that the polarization even increase faster when the volume is relaxed.}
		\label{fig:volume}
	\end{figure}

No qualitative difference is found in calculations with or without fixed volume for Sr$_3$Sn$_2$O$_7$.  The polarization strengths are calculated when volume is fixed and relaxed in the figure.~\ref{fig:volume}. When the volume is relaxed, the polarization is further enhanced under electron doping because the extra electron charge carriers expand the cell volume, which favors the polar modes. The tensile epitaxial strain has similar effects, as shown in the main text. 
	
	\subsection{Effect of Electronic Smearing Options}
\begin{figure}[!]
		\centering
		\begin{subfigure}[a]{\textwidth}
			\includegraphics[width=.7\textwidth]{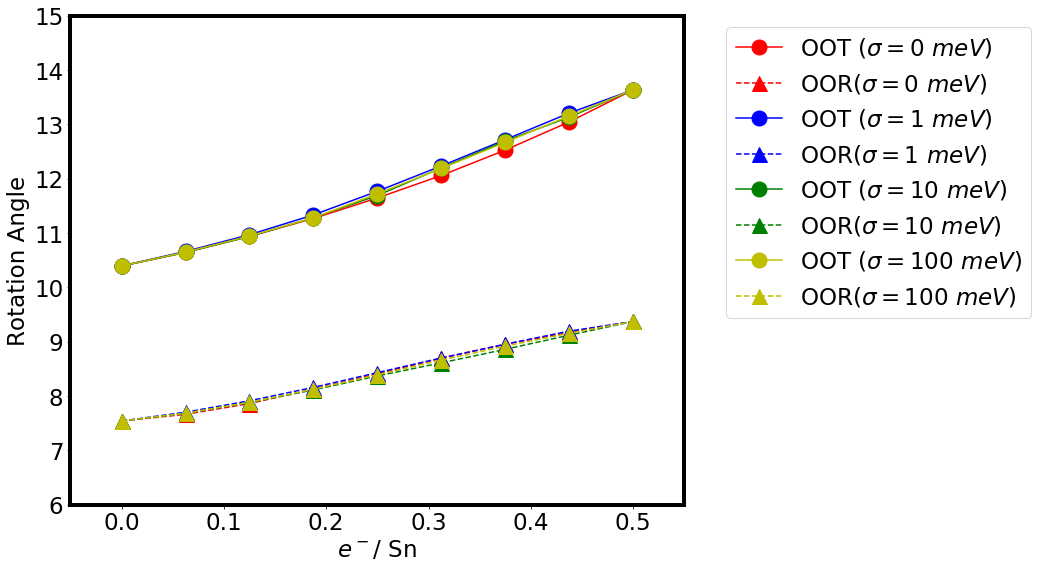}
			\caption{Rotation mode}
		\end{subfigure}
		
		\begin{subfigure}[b]{\textwidth}
			\includegraphics[width=.7\textwidth]{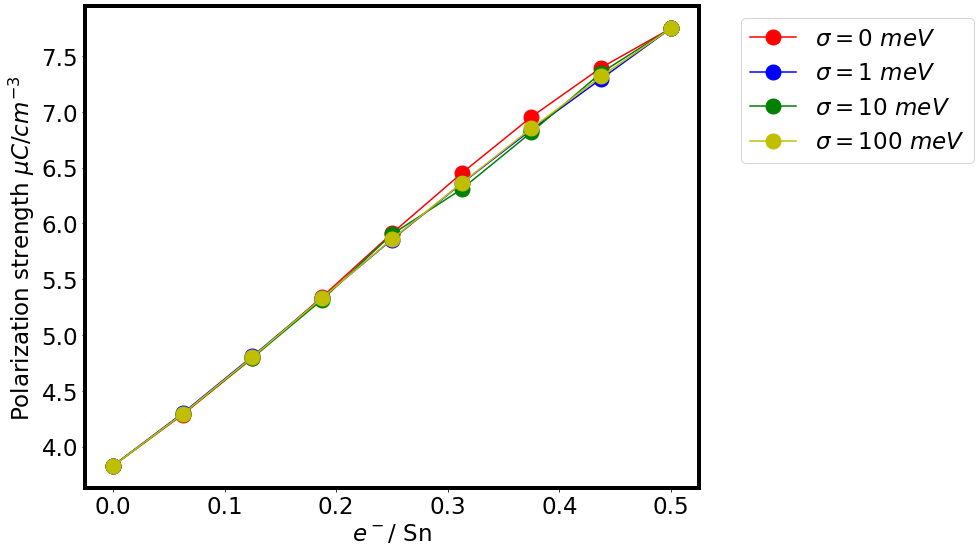}
			\caption{Polar mode}
		\end{subfigure}
		
		\caption{The oxygen octahedra rotation and net polarization of Sr$_3$Sn$_2$O$_7$ under doping, with different smearing parameters set in calculation. The bands are Gaussian-smeared using different $\sigma$ value. OOT stands for out-of-phase tilting ($X_3^-$ mode) and OOR stands for out-of-phase rotation($X_2^+$ mode). The volume is fixed in this calculation.}
		\label{fig:smearing}
	\end{figure}

	Electronic smearing has been applied to achieve better convergence of the electronic structure calculations of the metallic configurations. The smearing method and width can cause significant differences in the crystal structures and instabilities when there is Fermi surface nesting, but in the systems considered in this study, no such effect is observed for different width of Gaussian smearing. The results shown in Fig.~\ref{fig:smearing} show no systematic or significant difference for Sr$_3$Sn$_2$O$_7$. 

\subsection{{SrBa$_2$}Sn$_2$O$_7$}
	\subsubsection{Site-substitution}
	\begin{figure}
		\centering
		\includegraphics[width=.7\textwidth]{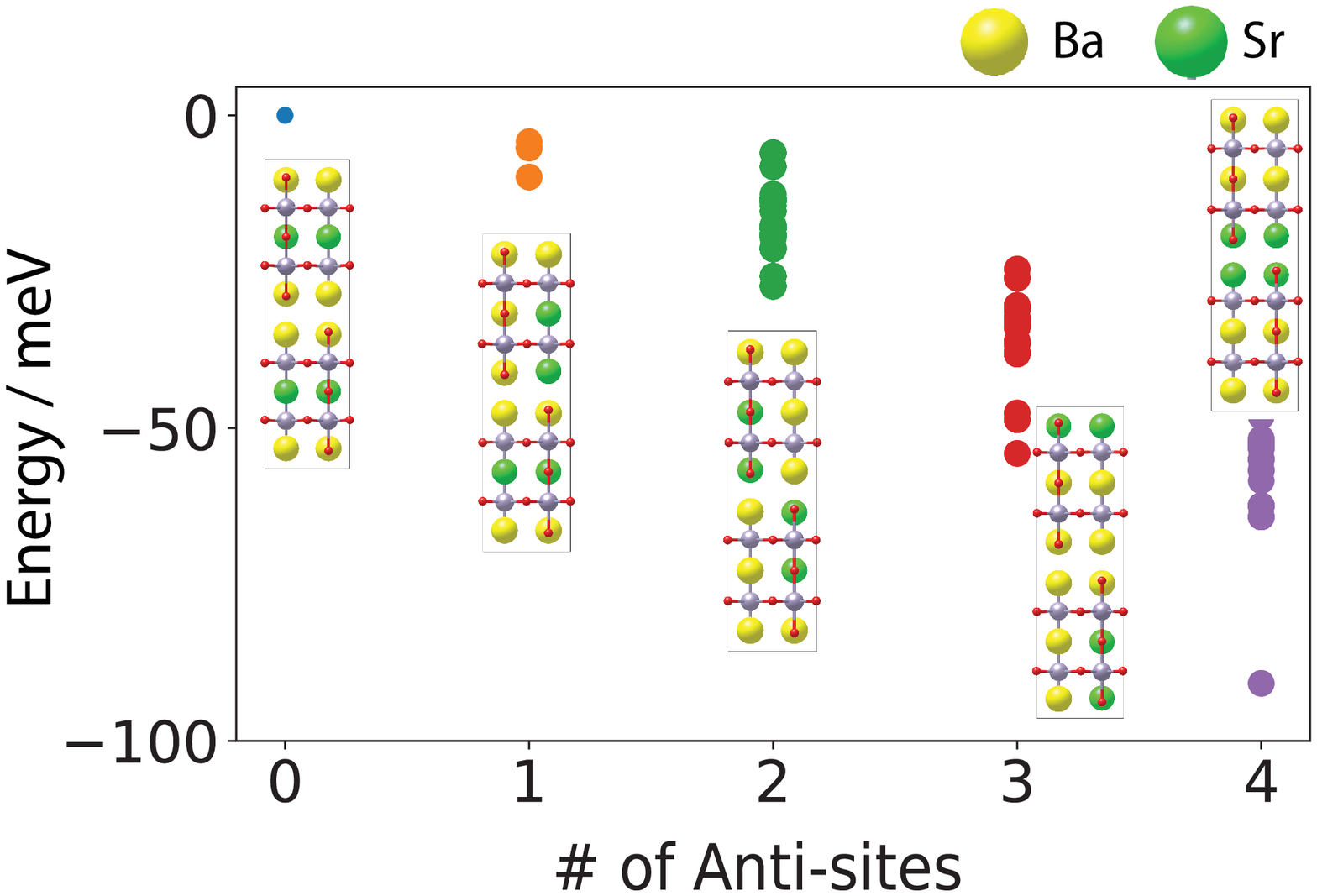}
		\caption{Energy as a function of the number of anti-sites in {SrBa$_{2}$}Sn$_2$O$_7$ compounds. The energy is plotted using dots and in scale of per Sn atom. The inner graphs are the crystal structures with different numbers of anti-sites in one conventional unit cell: 0-4 from left to right.}
		\label{fig:anti-site}
	\end{figure}
	
	The results in the maintext are for a particular layering pattern, or equivalently A-site cation order, for {SrBa$_2$}Sn$_2$O$_7$. In the pattern considered, the Ba atoms are on the Wyckoff position $e$ and Sr atoms are on the wyckoff position $b$, which preserves the $I4/mmm$ symmetry of {Sr$_3$}Sn$_2$O$_7$. While layer-by-layer growth methods such as molecular beam epitaxy makes growth of this structure possible in principle, it might not be the lowest energy structure in bulk. In this subsection , we show the results of calculations we performed in order to determine the lowest energy cation ordering pattern in {SrBa$_2$}Sn$_2$O$_7$. 

We performed calculations for all possible {SrBa$_2$}Sn$_2$O$_7$ structures with different A-site orders commensurate with the conventional tetragonal unit cell. Starting from the original $I4/mmm$ structure explained above, we define an 'anti-site' as one Ba atom swapping its position with a Sr atom in the conventional unit cell. The energies of structures with different numbers of anti-site defects are shown in fig.~\ref{fig:anti-site}. It can be seen that the more anti-sites the lower energy. The Ba atoms tend to migrate away from the vacuum layer because of the larger radius, which is a trend observed in other double-Ruddlesden-Popper compounds too. Accordingly, the smaller Sr atoms tend to migrate into the vacuum layer. This can be understood in connection with the significant rumpling present in the Ruddlesden-Popper compounds \cite{Birol2011}, which helps the smaller Sr ion have a more favorable coordination environment on the vacuum layer. 
	
\begin{figure}
	\centering
	\includegraphics[width=.7\textwidth]{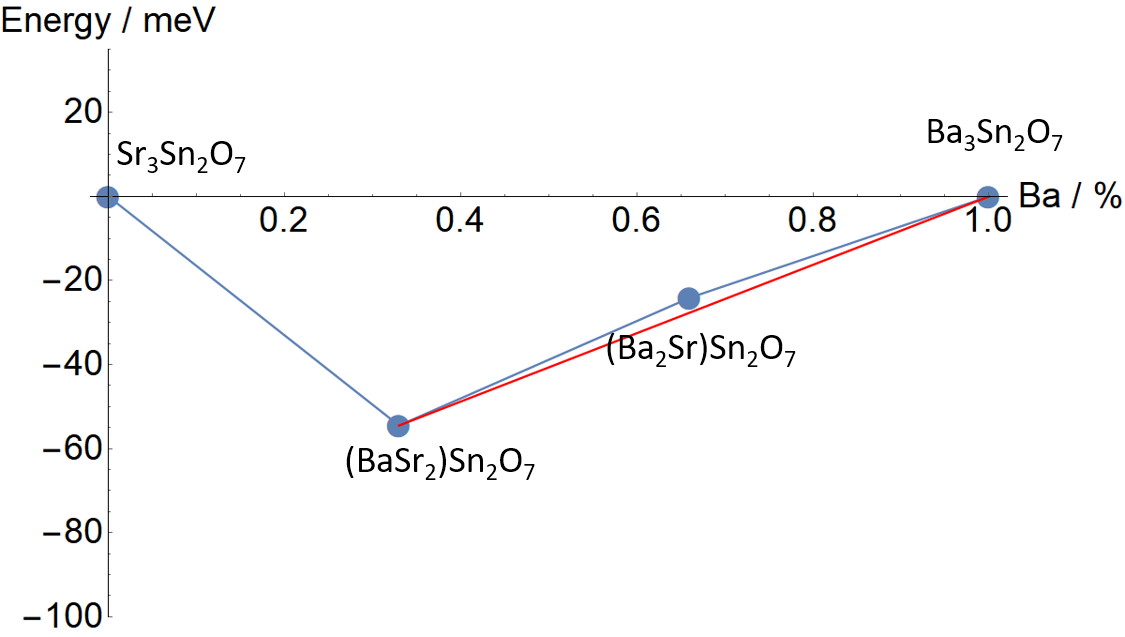}
	\caption{The convex hull for {Sr${_x}$,Ba$_{3-x}$}Sn$_2$O$_7$ compounds. The energy is in scale of per Sn atom. The {Sr${_x}$Ba$_{3-x}$}Sn$_2$O$_7$ is slightly over the convex hall which means it is thermodynamic unstable from our DFT calculations.}
	\label{fig:convex}
\end{figure}
	
In Fig.~\ref{fig:convex}, we show the convex hull which shows that Sr${_2}$BaSn$_2$O$_7$ is thermodynamically stable at zero temperature and pressure, but SrBa$_{2}$Sn$_2$O$_7$ is not, since it is above the convex hull. 

\subsubsection{Possible Structures of SrBa$_2$Sn$_2$O$_7$}
	
	\begin{table}[]
	\centering
	\begin{tabular}{|c|c|}
		\hline 
		Phonon mode          & Frequency / $cm^-1$  \\ \hline 
		$X_3^-$              & -110                \\
		$X_4^-$              & -101                \\
		$X_1^-$              & -57                 \\
		$X_2^+$              & -30                 \\
		$\Gamma_5^-$ (Polar) & 44             \\ \hline   
	\end{tabular}
	\caption{The frequency of unstable and polar phonon modes in {SrBa$_{2}$}Sn$_2$O$_7$ without anti-site defects.}
	\label{tab:phonon}
\end{table}

\begin{figure}[!]
		\centering
		\includegraphics[width=0.5\textwidth]{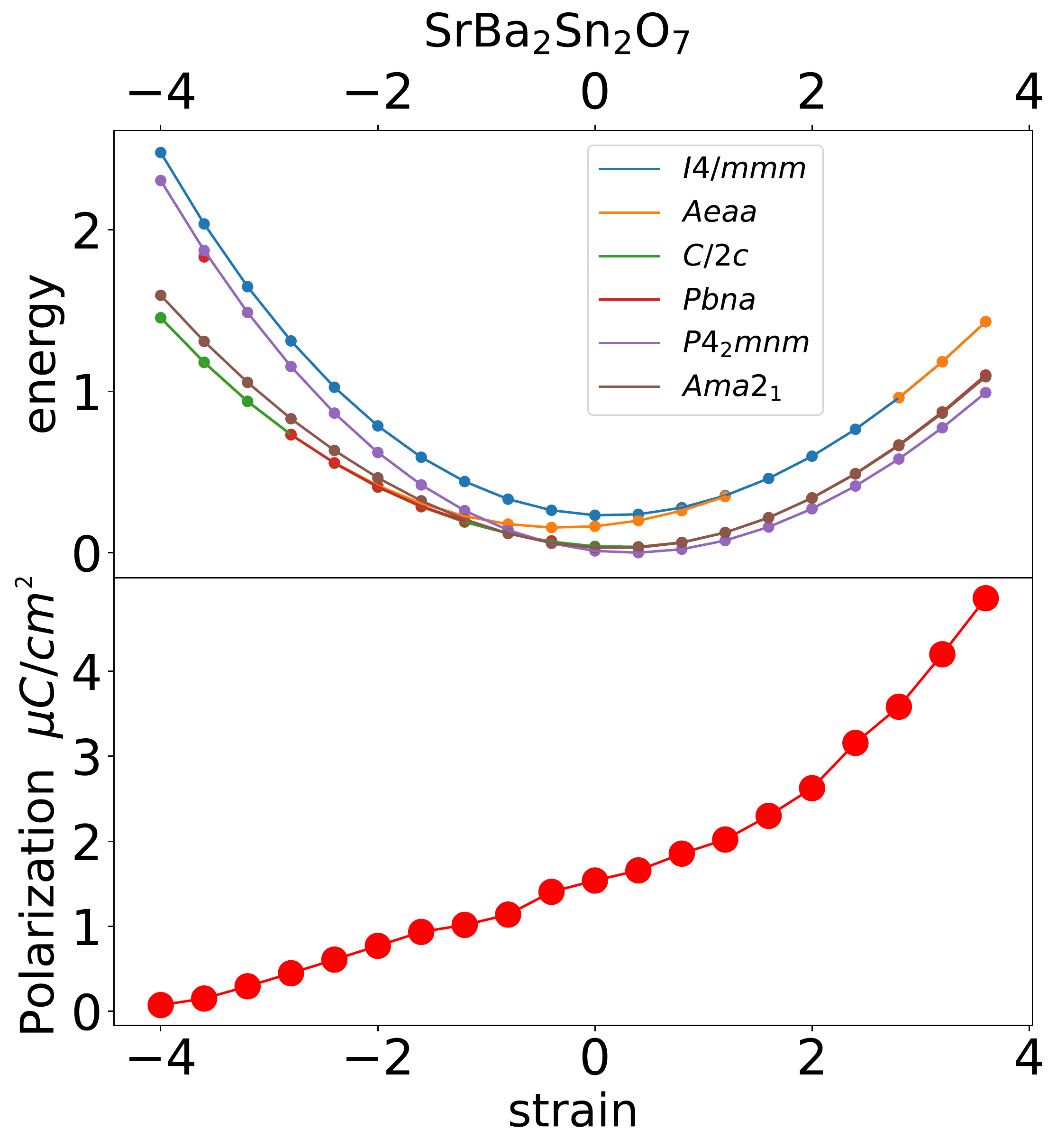}
		\caption{The energy of {SrBa$_{2}$}Sn$_2$O$_7$ compounds with no anti-site defects. Only the phases with energy close to ground state structure are shown here. The polarization plot is for the $Ama2_1$ phase, at all strain values.}
		\label{fig:energy}
	\end{figure}

	The reference high symmetry body-centered tetragonal phase of SrBa$_{2}$Sn$_2$O$_7$ has exactly the same symmetry as Sr$_{3}$Sn$_2$O$_7$, and hence, the other A$_3$B$_2$O$_7$ Ruddlesden-Popper hybrid improper ferroelectrics. Thus, the extensive group theoretical work on these compounds and their unstable phonons apply to SrBa$_{2}$Sn$_2$O$_7$ as well. The unstable phonons frequencies of SrBa$_{2}$Sn$_2$O$_7$, obtained from DFT, are shown in Table.~\ref{tab:phonon}. The $X_3^-$, $X_2^+$ and $X_1^-$ are all unstable as is the case in Sr$_{3}$Sn$_2$O$_7$. By combining these three modes and the other modes coupled with them at $\Gamma$ and $M$ points, a list of possible lower energy structures can be obtained (Table.~\ref{tab:trilinear}). All those phases have been considered as possible structural ground states of SrBa$_{2}$Sn$_2$O$_7$, and their energy are compared to get the phase diagram in the main text. Fig.~\ref{fig:energy} shows our calculation results of epitaxial strained SrBa$_{2}$Sn$_2$O$_7$ thin film with zero electron doping and its polarization strength. 

\subsubsection{Phase diagram of SrBa$_2$Sn$_2$O$_7$ with anti-sites}
	\begin{figure}[!]
	\centering
	\includegraphics[width=0.7\textwidth]{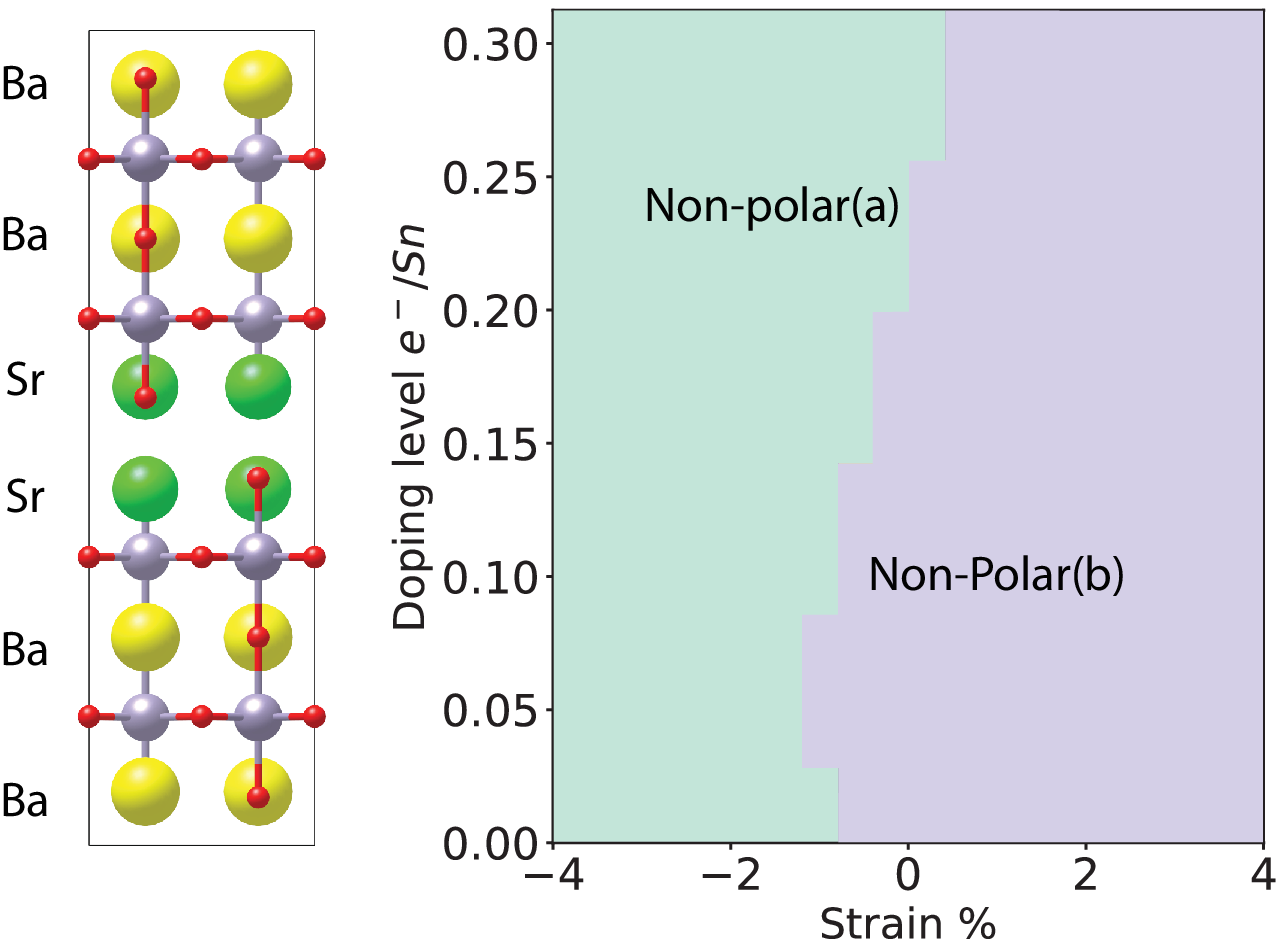}
	\caption{The phase diagram of {SrBa$_{2}$}Sn$_2$O$_7$ compounds with 4 anti-sites. There are two non-polar phases present in the phase diagram: non-polar(a) is $Aeaa$-like and non-polar(b) is $P4_2/mnm$-like.}
	\label{fig:phase_BSSO_4}
\end{figure}

In the main text, we show a phase diagram of {SrBa$_{2}$}Sn$_2$O$_7$ without anti-sites. However, as discussed earlier, {SrBa$_{2}$}Sn$_2$O$_7$ with 4 anti-sites has a lower energy. Here we perform another doping calculation on this new structures, which has the lowest energy among all {SrBa$_{2}$}Sn$_2$O$_7$ crystal structures we have studied. The results in Fig.~\ref{fig:phase_BSSO_4} indicate that only two phases are present and both of them are non-polar. The space group of these new structures are different from the ones of the structure without anti-site defects because the new cation ordering breaks a translation symmetry. Nevertheless, we used similar starting crystal structures as in Table.~\ref{tab:trilinear} to map out the possible phases, since the oxygen octahedra still have rotational instabilities in the similar environments.

\end{document}